\documentclass{jfm}
\usepackage{graphicx}
\usepackage{epstopdf, epsfig}

\usepackage{tcolorbox}

\usepackage{amsmath}
\usepackage{amsfonts}
\usepackage{amssymb}
\usepackage{graphicx}
\usepackage{longtable}
\usepackage{dcolumn}
\usepackage{bm}
\usepackage{epstopdf}
\usepackage{color}

\newcommand\bs{\boldsymbol}

\shorttitle{Morphodynamics of Active Nematic Fluid Surfaces}
\shortauthor{S.~C.~Al-Izzi and R.~G.~Morris}

\title{Morphodynamics of Active Nematic Fluid Surfaces}

\author{Sami C.~Al-Izzi\aff{1}\corresp{\email{s.al-izzi@unsw.edu.au}}
  \and Richard G.~Morris\aff{1}\corresp{\email{r.g.morris@unsw.edu.au}}}

\affiliation{\aff{1}School of Physics \& EMBL Australia Node in Single Molecule Science, School of Medical Sciences, University of New South Wales, Sydney 2052, Australia}
\begin{document}

\maketitle

\begin{abstract}
Morphodynamic equations governing the behaviour of active nematic fluids on deformable curved surfaces are constructed in the large deformation limit. Emphasis is placed on the formulation of objective rates that account for normal deformations whilst ensuring that tangential flows are Eulerian, and the use of the surface derivative (rather than the covariant derivative) in the nematic free energy, which elastically couples local order to out-of-plane bending of the surface. Focusing on surface geometry and its dynamical interplay with the hydrodynamics, several illustrative instabilities are then characterised. These include cases where the role of the Scriven-Love number and its nematic analogue are non-negligible, and where the active nematic forcing can be characterised by an analogue of the F\"{o}ppl-von-K\'{a}rm\'{a}n number. For the former, flows and changes to the nematic texture are coupled to surface geometry by viscous dissipation. This is shown to result in non-trivial relaxation dynamics for a nematic tube. For the latter, the nematic active forcing couples to the surface bending terms of the nematic free energy, resulting in extensile (active ruffling) and contractile (active pearling) instabilities in the tube shape, as well as active bend instabilities in the nematic texture. In comparison to the flat case, such bend instabilities now have a threshold set by the extrinsic curvature of the tube. Finally, we examine a topological defect located on an almost flat surface and show that there exists a steady state where a combination of defect elasticity, activity and non-negligible spin-connection drive a shape change in the surface.
\end{abstract}


\section{Introduction}
Two dimensional fluid surfaces with nematic ordering are relevant for the characterisation of a variety of biological and soft matter systems, ranging from liquid crystal shells \citep{Fernandez-Nieves07} to active composites of microtubules and kinesin motors at an oil-water interface \citep{Sanchez12}. Of particular current interest is the increasing experimental evidence that nematic-like ordering plays a vital role in tissue mechanics \citep{Saw17} and morphogenesis \citep{Maroudas-Sacks21,Vafa21}, where the interplay between topological defects and active forces control the morphology of protrusions and extrusions \citep{Metselaar19,Hoffmann21}.

Mathematically, the development of theory that couples hydrodynamics to surface geometry has lead to many interesting predictions on the role of viscous forces in the ordering and shaping of membranes in both isotropic \citep{Sahu20,Arroyo09,Hu07,Tchoufag22,Rangamani13,Steigmann99} and ordered fluids \citep{Napoli16,Nestler21}. Of particular note has been the inclusion of activity, leading to a variety of interesting morphodynamical phenomena and instabilities \citep{Salbreux17,Salbreux22,Bacher21,Khoromskaia21,Hoffmann21,Vafa21,Alert22,Bell22,Nestler21,Rank21} and even some attempts to construct rigorous shell theories of active materials \citep{Rocha22}. Much of this work has been driven by the desire to develop theories capable of describing the dynamics and morphogenesis of biological tissues and organisms \citep{Al-Izzi21,Julicher18}.

Despite such interest, there is still a lack of formal characterisation regarding how geometry couples to dissipatve and active forces in nematic fluid surfaces. To this end, we develop a closed system of dynamical equations for nematic fluid surfaces in a similar manner to the approach taken in \cite{Arroyo09} for isotropic fluid membranes. We construct normal and tangential dissipative and reactive forces for the system subject to a free energy that depends only on surface geometry, the order parameter and its derivatives. We focus specifically on nematic liquid crystals in the one-constant limit of the Frank free energy and derive the general form of the so-called molecular field and the resulting normal forces for surfaces of arbitrary geometry (and vector ordering). 
Particular emphasis is placed on the formulation of objective rates that account for normal deformations whilst ensuring that tangential flows are Eulerian, and the use of the surface derivative (rather than the covariant derivative) in the nematic free energy, which elastically couples local order to out of plane bending of the surface. Given the increased interest in active interfaces, we also include an active Q-tensor term in the tangential stress tensor. From these equations we identify and discuss three dimensionless numbers which characterise regimes where there is a non-trivial interplay between geometry, dissipation and active forces.

To highlight how these numbers govern morphodynamics, we focus on some simple concrete examples. In the first case, we consider a tube with no activity (and uniformly ordered nematic parameter as a ground-state) and derive the relaxation dynamics for perturbations in the shape and order parameter. We then consider the case of active tubes, where we show that under active forces there are three characteristic types of instability depending on the value and sign of the activity (extensile vs contractile). Two of these instabilities are in the tube shape: contractile forces give an active pearling-like instability whose lengthscale is set by the bulk viscous time-scale, whilst extensile forces (above a threshold) lead to a ruffling instability whose lengthscale is set by the active stress. In addition, we find an instability in the texture that acts to drive spontaneous bend. This occurs for extensile activity in rod-like nematics, and for contractile activity in disk-like nematics. Interestingly, although similar to known spontaneous bend instabilities in flat geometries, the effects of curvature mean that the active forcing must exceed a finite threshold even in the infinite system size. Finally, we examine steady states in shape driven by active forcing of an almost flat surface in the presence of a topological defect and show that this leads to protrusions at the defect whose morphology is controlled by the activity. Specifically, increases in contractility lead to increased protrusion, whilst increases in extensility act to suppress protrusions. 

\section{A Theory of Deformable Nematic Fluid Surfaces}
\begin{figure}
\center\includegraphics[width=0.6\textwidth]{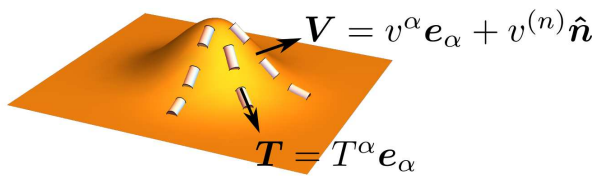}
\caption{\label{fig:schematic}Schematic of a curved fluid surface with orientational order given by the director field $\bs{T}=T^\alpha\bs{e}_\alpha$ which moves with a velocity in $\mathbb{R}^3$ given by the vector field $\bs{V} = v^\alpha\bs{e}_\alpha +v^{(n)}\bs{\hat{n}}$.}
\end{figure}

In this section, we construct a closed system of equations for the morphodynamics of nematic fluid surfaces (Fig.~\ref{fig:schematic}).  In general, certain conventions from differential geometry will be adopted, such as the use of comma ``$,$'' and semicolon ``$;$'' subscripts to denote partial and covariant differentiation, respectively. For the uninitiated, a brief non-rigorous introduction to such notation is given in Appendix \ref{sec:diffGeoApp}. For a more complete treatment, we refer the reader to \citet{Needham21} and/or \citet{Frankel11}.

Our starting point is a two-dimensional Riemannian manifold, $\mathcal{M}_t$, embedded in $\mathbb{R}^3$, and equipped with induced metric, $g_{\alpha\beta}$, and second fundamental form, $b_{\alpha\beta}$, where $\alpha, \beta \in \{1,2\}$ (and similarly for all Greek indices). The time-dependent position, $\bs{R}(u^\alpha(t),t)\in\mathbb{R}^3$, of each point $u\in\mathcal{M}_t$ implies a local  velocity
\begin{align}
\frac{\mathrm{d}\bs{R}}{\mathrm{d}t}= \bs{V}=v^\alpha\bs{e}_\alpha + v^{(n)}\bs{\hat{n}},\label{eq:V}
\end{align}
where the $\bs{e}_\alpha=\bs{R}_{,\alpha}$ are tangent to the surface at $\bs{R}$, and $\bs{\hat{n}}$ is the unit normal. To account for local order, each point is also associated with a vector $\bs{T}=T^\alpha \bs{e}_{\alpha}\in\mathcal{T}(\mathcal{M}_t)$, referred-to as the director, which lives in the tangent bundle of the manifold.

Throughout, the fluid is assumed to be in the low Reynolds number regime and therefore inertial forces are neglected. As a result, the condition of force balance can be decomposed in the following way
\begin{align}
    &\bs{f}^e+\bs{f}^d+ \bs{f}^{bs} + \bs{f}^c + \bs{f}^a= 0,
\end{align}
where $\bs{f}^e$ are elastic forces, $\bs{f}^d$ are dissipative forces, $\bs{f}^{bs}$ are broken symmetry forces, $\bs{f}^c$ are forces from constraints, and $\bs{f}^a$ are active forces.

We will also assume that the fluid is incompressible.  As a result, the only dynamical equation in our theory, in addition to the trivial (\ref{eq:V}), is for the director field $T^{\alpha}$ on the surface. We write this as
\begin{equation}
    \mathrm{D}_t T^\alpha = J^\alpha_{T} + J^\alpha_{c},
\end{equation}
where $\mathrm{D}_t$ is the objective rate (calculated in the next section), $J^\alpha_T$ are dissipative currents resulting from an Onsager expansion/thermodynamic flux-force relation, and $J^\alpha_c$ is the current associated with the constraint of unit director magnitude.

\subsection{Rate-of-deformation and vorticity tensors}
A tensor of vital importance to fluid mechanics is the rate-of-deformation, or strain-rate tensor, $d_{\alpha\beta}$. The rate-of-deformation tensor is given by half of the time derivative of the metric on $\mathcal{M}_t$ [more formally, half of the Lie-derivative with respect to the flow, $\bs{V}$ \citep{Arroyo09,Marsden94}]. If we take the time derivative of the induced metric, $g_{\alpha\beta}=\bs{e}_\alpha\cdot\bs{e}_\beta$ as defined in (\ref{eq:metricDef}), we find the following
\begin{equation}
\frac{\mathrm{d} g_{\alpha\beta}}{\mathrm{d}t} = \frac{\mathrm{d}\bs{e}_{\alpha}}{\mathrm{d}t}\cdot\bs{e}_{\beta}+ \frac{\mathrm{d}\bs{e}_{\beta}}{\mathrm{d}t}\cdot\bs{e}_{\alpha} = \bs{V}_{,\alpha}\cdot\bs{e}_{\beta} + \bs{e}_{\alpha}\cdot\bs{V}_{,\beta}\text{.}
\end{equation}

\noindent Substituting for $\bs{V}_{,\alpha}=\left(v^\beta{}_{;\alpha} - v^{(n)} b^{\beta}{}_{\alpha}\right)\bs{e}_{\beta} + \left(v^\beta b_{\alpha\beta} + v^{(n)}{}_{;\alpha}\right)\bs{\hat{n}}$ [see (\ref{eq:coordinateRate})] gives
\begin{equation}
d_{\alpha\beta}=\frac{1}{2}\frac{\mathrm{d} g_{\alpha\beta}}{\mathrm{d}t} = \frac{1}{2}\left(v_{\beta;\alpha} + v_{\alpha;\beta}\right) -  v^{(n)} b_{\alpha\beta}\text{.}
\end{equation}

\noindent Alternately, one can start from the expression for the rate-of-deformation tensor in $\mathbb{R}^3$ and project back onto the manifold $\mathcal{M}_{t}$ 
\begin{align}
d_{\alpha\beta}\bs{e}^\alpha\otimes\bs{e}^\beta=\left[\frac{1}{2}\bs{V}_{,\alpha}\otimes\bs{e}^{\alpha}+\frac{1}{2}\left(\bs{V}_{,\alpha}\otimes\bs{e}^{\alpha}\right)^{\text{T}}\right]^{\bullet}\text{,}
\end{align}
where $(\cdot)^{\bullet}$ denotes projection onto $\mathcal{M}_t$ (\textit{i.e.}~contracting with tangent basis $\bs{e}_{\beta}$) which a quick calculation verifies gives the same result. Note that $\bs{V}\in\mathbb{R}^3$ but has support only on $\mathcal{M}_t$ and that the formal derivation of this requires some subtle constructions such as extensions of vector fields \citep{Lee1997}.

The vorticity tensor can be derived in an equivalent way giving an identical result to the flat geometry version
\begin{equation}
\Omega_{\alpha\beta}=\frac{1}{2}\left(v_{\beta;\alpha}-v_{\alpha;\beta}\right)\text{,}
\end{equation}
which can be seen by noting that the terms of the asymmeterised deformation tensor proportional to $b_{\alpha\beta}$ vanish by symmetry under exchange of indices of the second fundamental form.

Here, we will consider incompressible fluids, the condition for which is that the rate-of-deformation tensor is traceless
\begin{equation}\label{eq:2Dincomp}
d^\alpha{}_\alpha = \nabla_\alpha v^\alpha - 2 H v^{(n)}=0\text{.}
\end{equation}
In the absence of normal flows, this reproduces the classic incompressibility condition, $\nabla_\alpha v^\alpha=0$.

\subsection{Objective rate of a vector field}
The subject of objective rates is contentious in continuum mechanics, especially where changing geometry is concerned \citep{Marsden94,Nitschke22}. The challenge, in this case, is to ensure that tangential flows are Eulerian--- {\it i.e.}, coordinates are not advected with the flow--- yet also properly account for out-of-plane surface dynamics, which are necessarily Lagrangian--- {\it i.e.}, coordinates {\it are} advected with surface movement \citep{Waxman84}. We adopt a pragmatic approach, and choose a projected co-rotational rate. That is, we take the co-rotational derivative in $\mathbb{R}^3$ and project down onto the surface so as to recover the co-rotational objective rates for fixed-curved and flat surfaces, while still accounting for Lagrangian deformations of the surface in the normal direction.

For a vector field $\bs{T}$, which is confined to the tangent bundle of $\mathcal{M}_{t}$--- \textit{i.e.}~$\bs{T}=T^\alpha\bs{e}_\alpha\in \mathcal{T}\left(\mathcal{M}_t\right)$--- this takes the form
\begin{equation}
\text{D}_t\bs{T}^{\bullet} =\left[\frac{\mathrm{d}\bs{T}}{\mathrm{d}t} + \Omega\cdot\bs{T}\right]^{\bullet} = \left[\partial_t\bs{T} +v^\alpha\left(\bs{T}\right)_{,\alpha} + \Omega\cdot\bs{T}\right]^{\bullet}\text{.}
\end{equation}

\noindent The first term is given by
\begin{align}
\partial_t\bs{T} &= \partial_{t}\left(T^\alpha\bs{e}_\alpha\right) = \partial_tT^\alpha \bs{e}_\alpha + T^\alpha \partial_t\bs{e}_\alpha\nonumber\\
& = \left(\partial_tT^\alpha - v^{(n)} T^\beta b_{\beta}{}^{\alpha}\right)\bs{e}_\alpha +  T^\alpha v^{(n)}{}_{;\alpha}\bs{\hat{n}}\text{,}
\end{align}
where we have only included contributions from changing the basis vectors due to normal velocities, since we are in an Eulerian frame tangentially and therefore tangential velocities do not advect coordinates. The second term is given by
\begin{align}
v^{\alpha}\nabla_\alpha\left(\bs{T}\right) & = v^\alpha\left(T^\beta \bs{e}_\beta\right)_{,\alpha} = v^\alpha T^\beta{}_{;\alpha}\bs{e}_\beta + T^\beta v^{\alpha} b_{\alpha\beta}\bs{\hat{n}}\text{,}
\end{align}
and finally the third is simply given by
\begin{equation}
\Omega\cdot\bs{T}=\Omega^{\alpha}{}_{\beta}T^\beta\bs{e}_{\alpha}\text{.}
\end{equation}

\noindent Putting this all together and projecting back onto the manifold gives
\begin{align}
\text{D}_tT^\alpha = &\partial_tT^\alpha + v^\beta T^\alpha{}_{;\beta} + \Omega^{\alpha}{}_{\beta}T^\beta - v^{(n)} T^\beta b_{\beta}{}^{\alpha}\text{,}
\end{align}

\noindent which is identical to the equivalent expression in flat geometry, with the exception of the final term.

To understand the origin of this term, consider a tube in cylindrical polar coordinates that is expanding outwards with velocity $v^{(n)}=\dot{R}$, where $R$ is the radius. There is no tangential flow, and the director field is assumed to be of constant magnitude in the $\bs{e}_{\theta}$ direction ({\it i.e.},  $\bs{T}=T\bs{e}_{\theta}$). In this case, the second fundamental form has one non-zero component in the $\bs{e}_\theta$ direction. As a result, as $R$ increases, so does the size of the basis vector $\bs{e}_\theta$, and we require that $\mathrm{D}_tT^{\theta} = \partial_t T + \dot{R} T/R=0$ for the objective rate to be zero (fixed size for $\bs{T}$), which gives $T=T_0\exp(-\dot{R}/R)$.

\subsection{Elastic forces}
In general, we will consider surfaces with free energies of the following form
\begin{equation}
\mathcal{F}=\int_{\mathcal{M}_t}F\left(g_{\alpha\beta},b_{\alpha\beta},T_{\alpha}\right)\mathrm{d}A\text{,}
\end{equation} 
where $\mathrm{d}A=\sqrt{|g|}\mathrm{d}x^1\mathrm{d}x^2$ is the area element on the manifold. Whilst other free energies fit into this class, for example the tilt field of a lipid monolayer \citep{Terzi17,Selinger96}, we will focus on the concrete example of the free energy of a nematic liquid crystal. Here, we will make use of standard descriptions of liquid crystals to define the molecular field as
\begin{equation}
    h_\alpha = - \frac{\delta \mathcal{F}}{\delta T^\alpha}\text{,}
\end{equation}
which we will make use of in order to derive the hydrodynamics of the ordered surface. The normal force given by varying the surface shape is then given by
\begin{equation}
    \bs{f}_{\text{e}}=f^\alpha_{\text{e}}\bs{e}_\alpha + f^{(n)}_{\text{e}}\bs{\hat{n}} = -\frac{\delta \mathcal{F}}{\delta \bs{R}}\mathrm{.}
\end{equation}

In the case of nematic liquid crystals, the one-constant approximation to the Frank free energy \citep{Frank58,DeGennes93,Napoli12,Napoli16} is given by the following
\begin{equation}
\mathcal{F}^{\text{LC}} = \int_{\mathcal{M}_t} \frac{\mathcal{K}}{2}|\tilde{\nabla}_{s}\bs{T}|^2\mathrm{d}A \text{,}
\end{equation}
where $\bs{T} = T^\beta \bs{e}_\beta$ is the nematic order parameter (we choose to use $T^\beta$ here rather than the usual $n^\beta$ in the nematic liquid crystal literature in order to avoid confusion with the surface normal vector $\bs{\hat{n}}$) on the tangent bundle $\mathcal{T}(\mathcal{M}_t)$ and $\tilde{\nabla}_s$ is the surface derivative (not the covariant derivative) \citep{Napoli12}. The reason for our choice of the surface derivative, as opposed to the covariant derivative, is so as to correctly account for the bend due to extrinsic curvature. This is likely an important coupling for true active nematic surfaces, but is perhaps less relevant for nematic-like material surfaces such as epithelial tissues. The surface derivative can be written in terms of the covariant derivative and second fundamental form as follows
\begin{equation}
\tilde\nabla_s \bs{T} = \nabla_\alpha T^\beta \bs{e}^\alpha\otimes\bs{e}_\beta + T^\beta b_{\alpha\beta} \bs{e}^\alpha\otimes\hat{\bs{n}}\text{,}
\end{equation}
and thus the full free energy is given by
\begin{equation}
\mathcal{F}^{\text{LC}} = \int_{\mathcal{M}_t} \frac{\mathcal{K}}{2}\left[\nabla_\alpha T^\beta \nabla^\alpha T_\beta + T^\beta T_\gamma b_{\alpha\beta}b^{\alpha\gamma}\right]\mathrm{d}A\text{.}
\end{equation}

We will now compute the associated functional derivatives in the free energy rate ($\dot{\mathcal{F}}$). The functional derivative with respect to $T^\alpha$ gives the negative of the molecular field $h_\alpha$ as follows
\begin{align}\label{eq:h_alpha}
\frac{\delta \mathcal{F}^{\text{LC}}}{\delta T^\alpha} = - h_\alpha = -\mathcal{K} \left[ \Delta T_\alpha + \left(K\delta_\alpha^\gamma - 2H b_\alpha{}^\gamma\right)T_\gamma \right]\text{.}
\end{align}

\noindent We will make use of the rate formulas calculated in Appendix \ref{sec:diffGeoApp} to perform functional variations of the form
\begin{align*}
&\delta g_{\alpha\beta} = \delta t \frac{\mathrm{d}}{\mathrm{d}t}\left( g_{\alpha\beta}\right)+O(\delta t^2)\text{,}\\
&\delta b_{\alpha\beta} = \delta t \frac{\mathrm{d}}{\mathrm{d}t}\left( b_{\alpha\beta}\right)+O(\delta t^2)\text{,}\\
&\delta \Gamma^{\gamma}_{\alpha\beta} = \delta t \frac{\mathrm{d}}{\mathrm{d}t}\left( \Gamma^\gamma_{\alpha\beta}\right)+O(\delta t^2)\text{.}
\end{align*}
Using this we find
\begin{align}
&\delta\mathcal{F} = \int_{\mathcal{M}_t} \frac{\mathcal{K}}{2} \Bigg[ 2 \delta(\Gamma^\beta_{\alpha\mu}) T^\mu g^{\alpha\gamma} g_{\beta\delta}  T^{\delta}{}_{;\gamma} + T^\beta{}_{;\alpha} \delta(g^{\alpha\gamma}) g_{\beta\delta} T^{\delta}{}_{;\gamma} + T^\beta{}_{;\alpha} g^{\alpha\gamma} \delta(g_{\beta\delta}) T^{\delta}{}_{;\gamma}\nonumber\\
&+ 2 \delta (b_{\alpha\beta}) g^{\alpha\delta}b_{\delta\gamma} T^\gamma T^\beta + T^\beta T^\gamma b_{\alpha\beta}b_{\delta\gamma} \delta(g^{\alpha\delta}) -2H v^{(n)} \delta t\bigg( T^\beta{}_{;\alpha} T_\beta{}^{;\alpha} + T^\beta T_\gamma b_{\alpha\beta} b^{\alpha\gamma}\bigg)\Bigg]\mathrm{d}A\text{.}
\end{align}
Focusing purely on the normal variation we find
\begin{align}
&= \int_{\mathcal{M}_t}\mathcal{K} \delta t\Big\{ \big[ (v^{(n)} b_{\alpha}{}^\gamma)_{;\beta} - (v^{(n)}b_{\alpha\beta})^{;\gamma} - (v^{(n)}b^\gamma{}_\beta)_{;\alpha}\big]T^\alpha T^\beta{}_{;\gamma} + T^\beta{}_{;\alpha}T_{\beta;\gamma}b^{\alpha\gamma}v^{(n)} \nonumber\\
&- T^\beta{}_{;\alpha} T^{\gamma;\alpha}b_{\gamma\beta}v^{(n)} +(b^\alpha{}_\gamma T^\gamma T^\beta)_{;\beta\alpha}v^{(n)} - T^\gamma T^\beta \big(2H b_{\alpha\beta}b^\alpha{}_\gamma - K b_{\beta\gamma} \big)v^{(n)}\nonumber\\
&+ T^\beta T^\gamma b_{\alpha\beta} b_{\delta\gamma}b^{\alpha\delta}v^{(n)}- H \big[ T^\beta{}_{;\alpha} T_\beta{}^{;\alpha} + T^\beta T_\gamma b_{\alpha\beta} b^{\alpha\gamma}\big]v^{(n)}\Big\}\mathrm{d}A\text{.}
\end{align}
Notice that the first terms from the variation of the covariant derivative include derivatives in the normal velocity. Integrating these by parts and pushing terms to the boundary we can find the following functional variation in the bulk of the surface
\begin{align}
\delta\mathcal{F} =& \int_{\mathcal{M}_t}\mathcal{K} v^{(n)}\delta t\Big\{  b_{\alpha\beta}(T^\alpha T^\beta{}_{;\gamma})^{;\gamma} + b^\gamma{}_\beta(T^\alpha T^\beta{}_{;\gamma})_{;\alpha}- b_{\alpha}{}^\gamma (T^\alpha T^\beta{}_{;\gamma})_{;\beta} + T^\beta{}_{;\alpha}T_{\beta;\gamma}b^{\alpha\gamma}\nonumber\\
&- T^\beta{}_{;\alpha} T^{\gamma;\alpha}b_{\gamma\beta} + (b^\alpha{}_\gamma T^\gamma T^\beta)_{;\beta\alpha} - T^\gamma T^\beta \big(2H b_{\alpha\beta}b^\alpha{}_\gamma - K b_{\beta\gamma} \big) + T^\beta T^\gamma b_{\alpha\beta} b_{\delta\gamma}b^{\alpha\delta}\nonumber\\
 &- H \big[ T^\beta{}_{;\alpha} T_\beta{}^{;\alpha} + T^\beta T_\gamma b_{\alpha\beta} b^{\alpha\gamma}\big]\Big\}\mathrm{d}A + \text{Boundary Terms}\text{.}
\end{align}
From this, and noting that the second and third terms cancel due to symmetry of $T^\alpha T^\beta$ and $b_{\alpha\beta}$, we find the forces in the normal direction from the free energy as the following
\begin{align}\label{eq:normalLCforce}
&f^{(n)}_{\text{e}}=-\mathcal{K}\Big\{ b_{\alpha\beta}(T^\alpha T^\beta{}_{;\gamma})^{;\gamma} + T^\beta{}_{;\alpha}\left(T_{\beta;\gamma}b^{\alpha\gamma} - T^{\gamma;\alpha}b_{\gamma\beta}\right)+(b^\alpha{}_\gamma T^\gamma T^\beta)_{;\beta\alpha}\nonumber\\
&- T^\gamma T^\beta \big(2H b_{\alpha\beta}b^\alpha{}_\gamma - K b_{\beta\gamma} \big) + T^\beta T^\gamma b_{\alpha\beta} b_{\delta\gamma}b^{\alpha\delta} - H \big[ T^\beta{}_{;\alpha} T_\beta{}^{;\alpha} + T^\beta T_\gamma b_{\alpha\beta} b^{\alpha\gamma}\big]\Big\}\text{.}
\end{align}
Note that the third term gives forces which go like the second derivative of the curvature tensor, but traced out along the directions of the nematic order parameter. Such a term is essentially an anisotropic bending rigidity that would not be found in the case where the covariant derivative, rather than surface derivative, was used in the free energy \citep{Napoli12,Santiago19,Santiago20}. This gives significantly different equations to many of those currently analysed in the literature; both equilibrium and dynamical \citep{Frank08,Salbreux22,Hoffmann21,Santiago18}.

Similarly for the tangential variations we find the following elastic force
\begin{align}
    f^{\alpha}_{\text{e}} = -\frac{\mathcal{K}}{2} \bigg\{&  \left(T^\beta T^{\alpha;\gamma}\right)_{;\beta\gamma} + \left(T_{\beta;\gamma} T^{\beta;\alpha}\right)^{;\gamma} - \left(T_{\beta;\gamma} T^{\alpha;\gamma}\right)^{;\beta} - 2 \left(b^\alpha{}_\beta b^\gamma{}_\mu T^{\mu\beta}\right)_{;\gamma} \nonumber\\
    &- 2 \left( b^\gamma{}_\mu T^{\mu\beta}\right)_{;\beta} b^\alpha{}_\gamma + \left(T^\beta T^\gamma b_{\mu\beta} b^\alpha{}_{\gamma}\right)^{;\mu}\bigg\}\text{.}
\end{align}
This is essentially a generalised version of the Ericksen force \citep{DeGennes93}. For simplicity we will neglect this tangential component in our later examples as it leaves the phenomenology of the equations unchanged. This approximation is taken in many exact calculations in active hydrodynamics \citep{Hoffmann21,Giomi14,Kruse04,Edwards09}, however in principle this term should be included in full non-linear solutions.

\subsection{Surface Tension and Director Constraints}
We consider two constraints upon our system whose contributions to the dynamical equations can generically be derived from a Rayleigh dissipation functional \citep{DeGennes93,Doi11}.  However, we avoid a full variational treatment here for the sake of brevity and instead simply state the contributions whose derivation can be found elsewhere \citep{DeGennes93,Arroyo09,Marchetti13}.

The first constraint is that of constant area, or incompressibility, which is expressed by (\ref{eq:2Dincomp}). The corresponding Lagrange multiplier is the surface tension, $\zeta$, which acts isotropically via stresses $\sigma^{(st)}_{\alpha\beta}=\zeta g_{\alpha\beta}$. Taking the surface divergence of $\sigma^{(st)}$ gives the following  contribution to the force balance condition
\begin{equation}
\bs{f}^\text{c} = \nabla^\alpha \zeta\bs{e}_\alpha + 2H \zeta \bs{\hat{n}}\text{.} 
\end{equation}

The second constraint is to fix the magnitude of the nematic director \citep{DeGennes93}. This is achieved by imposing 
\begin{equation}
    T_\alpha \mathrm{D}_t T^\alpha = 0\text{,}
\end{equation}
which implies $T^\alpha T_\alpha=\text{constant}$, with the constant specified by the initial conditions on $T^\alpha$ \citep{Kruse05}.  The associated Lagrange multiplier we label by $\lambda$, and the corresponding contribution to the director current \citep{Prost15} is just
\begin{equation}\label{eq:constraintCurrent}
    J^\alpha_{c} = -\lambda T^\alpha\text{.}
\end{equation}
This term can be viewed from a variational perspective as either a constraint on the Rayleigh dissipation functional of the form, $\int \lambda T_\alpha \mathrm{D}_t T^\alpha\mathrm{d}A$, or a contribution to the molecular field from a constraint on the free energy of the form, $\int\frac{1}{2}\lambda \gamma T_\alpha T^\alpha \mathrm{d}A$ \citep{DeGennes93}.

\subsection{Constitutive Relation and Dissipative Forces}
The viscous stress tensor in the tangent bundle is given as an Onsager expansion \citep{Doi11,DeGennes93} in the thermodynamic fluxes. At lowest non-trivial order, it is given by
\begin{align}
\sigma_{\alpha\beta} = 2 \eta d_{\alpha\beta} + \frac{\nu}{2}\left(T_\alpha h_\beta + T_\beta h_\alpha\right)\text{,}
\end{align}
where $h_\alpha$ is the molecular field (\ref{eq:h_alpha}), $d_{\alpha\beta}$ the rate-of-deformation tensor, $\eta$ is the shear viscosity and $\nu$ the spin connection coefficients of the vector field (essentially the ratio of anisotropic viscosity to isotropic viscosity \citep{DeGennes93}). Here stable solutions require $|\nu|>1$, with $\nu<-1$ corresponding to rod-like nematics and $\nu>1$ to disk-like nematics \citep{Edwards09,Marchetti13}. Note that, as we only consider incompressible fluids, we discard the bulk viscosity terms and associated spin connections \citep{DeGennes93}. The dissipative forces on our surface are then given as
\begin{equation}
\bs{f}^\text{d} = \nabla_\beta \sigma^{\alpha\beta}\bs{e}_\alpha + \sigma^{\alpha\beta}b_{\alpha\beta} \bs{\hat{n}}\text{.} 
\end{equation}

For the terms in the tangent bundle we find the following
\begin{align}
\nabla_\beta\sigma^{\alpha\beta} = 2 \eta d^{\alpha\beta}{}_{;\beta} + \frac{\nu}{2}\left(T^\alpha h^\beta + T^\beta h^\alpha\right)_{;\beta}\text{,}
\end{align}
where we can unpack the first term to
\begin{align}
 2\eta d^{\alpha\beta}{}_{;\beta} = \eta \left(v^{\alpha;\beta} + v^{\beta;\alpha} - 2 v^{(n)}b^{\alpha\beta}\right)_{;\beta} = \eta\Delta v^\alpha + \eta v^\beta{}_{;\alpha\beta} - 2\eta \left(v^{(n)}b^{\alpha\beta}\right)_{;\beta}\text{,}
\end{align}
where $\Delta=\nabla_\alpha\nabla^\alpha$ is the Bochner Laplacian. By making use of the Codazzi equation, (\ref{eq:codazzi}), and contracted Gauss Equation, (\ref{eq:contractedGauss2D}) we find
\begin{align}
2\eta d^{\alpha\beta}{}_{;\beta} = \eta\bigg[ \Delta v^\alpha - 2\left(v^{(n)}H\right)^{;\alpha} + K v^{\alpha} + 2\left(2H g^{\alpha\beta} - b^{\alpha\beta}\right)v^{(n)}{}_{;\beta}\bigg]\text{.}
\end{align}
Thus the tangential forces are given by
\begin{align}
\sigma^{\alpha\beta}{}_{;\beta} = & \frac{\nu}{2}\left(T^\alpha h^\beta + T^\beta h^\alpha\right)_{;\beta}\nonumber\\
& +\eta\bigg[ \Delta v^\alpha - 2\left(v^{(n)}H\right)^{;\alpha} + K v^{\alpha} + 2\left(2H g^{\alpha\beta} - b^{\alpha\beta}\right)v^{(n)}{}_{;\beta}\bigg] \text{.}
\end{align}

The normal forces are given by contracting the stress tensor with the second fundamental form
\begin{align}
 \sigma^{\alpha\beta}b_{\alpha\beta} =  2\eta b^{\alpha\beta}v_{\beta;\alpha} - 4\eta\left(2 H^2 - K\right)v^{(n)}  + \nu T^{\alpha}h^{\beta}b_{\alpha\beta}\text{.} 
\end{align}

\subsection{Broken Symmetry Terms}
We can also include broken symmetry terms in the stress tensor which contribute nothing to the overall dissipation as their dynamics are associated with Goldstone modes in the system, hence the name ``Broken Symmetry'' \citep{DeGennes93,Kruse05}. For vector fields these contribute an anti-symmetric part to the stress tensor of the following form
\begin{equation}
\sigma^{(bs)}_{\alpha\beta} = \frac{1}{2}\left(T_\alpha h_\beta - T_\beta h_\alpha\right)\text{.}
\end{equation}
Note that these terms only contribute to the tangential forces as they vanish in the normal due to the symmetry of the second fundamental form upon index relabelling. The broken symmetry forces are thus given by
\begin{equation}
    \bs{f}^{bs} = \nabla_\beta \sigma^{(bs)\alpha\beta}\bs{e}_\alpha + \sigma^{(bs)\alpha\beta}b_{\alpha\beta} \bs{\hat{n}} = \frac{1}{2}\left(T^\alpha h^\beta - T^\beta h^\alpha\right)_{;\beta}\bs{e}_\alpha\text{.}
\end{equation}

\subsection{Director Terms}
Expanding out in an Onsager fashion for the director rate in terms of nematic relaxation, we find the following couplings
\begin{equation}
    J^{\alpha}_{T} = \gamma^{-1} h^\alpha -\nu d^\alpha{}_{\beta}T^\beta\text{,}
\end{equation}
which gives terms in a similar manner to the flat case \citep{Kruse05}. Note that the constant $\nu$ is the same as for the director terms in the stress tensor by the Onsager reciprocal relations. Adding this together with the constraint current, (\ref{eq:constraintCurrent}), gives the full Onsager expansion for the director rate as
\begin{equation}
    \mathrm{D}_t T^\alpha =\gamma^{-1} h^\alpha - \nu d^\alpha{}_\beta T^\beta - \lambda T^\alpha\text{.}
\end{equation}

\subsection{Active Terms}
We consider a simple extension to the tangential stress tensor that represent the dipolar-like active forces that are characteristic of many active nematic systems \citep{Sanchez12,Marchetti13}. For a detailed introduction to active matter we refer the reader to the following reviews \citep{Marchetti13,Ramaswamy17} and for active liquid crystals in particular to \citep{Doostmohammadi18}. 

Active liquid crystals have generated significant interest in recent years due to several experimental realisations \citep{Sanchez12,Keber14} which have revealed the fascinating interplay between active stresses and geometry of the director, particularly in the coupling of geometry of the surfaces to the dynamics of topological defects \citep{Giomi14,Pearce19,Pearce20,Khoromskaia17}. In addition such systems have been suggested as minimal models for tissues \citep{Blanch-Mercader21a,Blanch-Mercader21b} and the actomyosin cell cortex \citep{Kruse05}.

Here, we consider the standard form for the active nematic stress tensor:
\begin{equation}\label{eq:activeStress}
\sigma^{\text{(a)}}_{\alpha\beta} = \Sigma Q_{\alpha\beta} = \Sigma \left(T_\alpha T_\beta - \frac{1}{2} g_{\alpha\beta}\right)\text{.}
\end{equation}
For $\Sigma>0$ such stresses are contractile in nature and for $\Sigma<0$ they are extensile.
This form represents the simplest form of anisotropic active stress, however more complicated couplings are possible \citep{Salbreux22,Salbreux17,Naganathan14}. The active forces from (\ref{eq:activeStress}) are given by
\begin{align}
    \bs{f}^a = \sigma^{\text{(a)}\alpha\beta}{}_{;\beta} \bs{e}_{\alpha}+  \sigma^{\text{(a)}\alpha\beta}b_{\alpha\beta}\bs{\hat{n}}\text{.}
\end{align}

\subsection{Full Dynamical Equations}
Putting together all our force balance terms, constraints and polarisation currents gives the following system, which are essentially ordered fluid versions of the equations derived for isotropic fluid membranes in \cite{Arroyo09}.
\begin{align}
&\text{Polarisation dynamics:}\nonumber\\
&\mathrm{D}_t T^\alpha =\gamma^{-1} h^\alpha - \nu d^\alpha{}_\beta T^\beta - \lambda T^\alpha\text{,}\\
&\text{Tangential force balance:}\nonumber\\
& \frac{1}{2}\left(T^\alpha h^\beta - T^\beta h^\alpha\right)_{;\beta} +\eta\bigg[ \Delta v^\alpha - 2\left(v^{(n)}H\right)^{;\alpha} + K v^{\alpha} + 2\left(2H g^{\alpha\beta} - b^{\alpha\beta}\right)v^{(n)}{}_{;\beta}\bigg]\nonumber\\
&\quad\quad + \frac{\nu}{2}\left(T^\alpha h^\beta + T^\beta h^\alpha\right)_{;\beta} + f^\alpha_{\text{e}} + \zeta^{;\alpha}  + \sigma^{\text{(a)}\alpha\beta}{}_{;\beta}=0\text{,}\\
&\text{Normal force balance:}\nonumber\\
&2\eta b^{\alpha\beta}v_{\beta;\alpha} - 4\eta\left(2 H^2 - K\right)v^{(n)} + \nu T^{\alpha}h^{\beta}b_{\alpha\beta} + 2 H\zeta + f^{(n)}_{\text{e}} + \sigma^{\text{(a)}\alpha\beta}b_{\alpha\beta}=0\text{,}\\
&\text{Constraint equations (constant director and incompressibility):}\nonumber\\
& T_\alpha\mathrm{D}_t T^\alpha=0;\quad v^\alpha{}_{;\alpha}-2H v^{(n)}=0\text{,}\\
&\text{Closure between surface velocity and geometry:}\nonumber\\
& \bs{V}= \frac{\mathrm{d}\bs{R}}{\mathrm{d}t}\text{.}
\end{align}

\subsection{Dimensionless Numbers}
There is an interesting point to make regarding the normal force balance equation. It contains the term $\eta b^{\alpha\beta}v_{\beta;\alpha} $, which describes the curvature induced coupling between viscous forces and shape. Such couplings have been discussed extensively for isotropic fluid membranes \citep{Sahu20,Al-Izzi20}. These phenomena can be understood in terms of the dimensionless Scriven-Love number \citep{Scriven60}
\begin{equation}
    \text{SL} = \frac{O(\eta b^{\alpha\beta}v_{\beta;\alpha})}{O(\text{Bending Forces})} = \frac{\eta V L}{\mathcal{K}}\text{,}
\end{equation}
where $V$ and $L$ are characteristic velocities and length-scales and we have assumed that $\text{Bending Forces}\sim \mathcal{K} L^{-3}$ (\textit{i.e.}, like the bending energy and Laplacian of the mean curvature).

In addition, however, there is a new anisotropic coupling due to terms $\sim \nu T^{\alpha}h^{\beta}b_{\alpha\beta}$ which, when compared to bending forces, could be thought of as a ``Liquid Crystalline Scriven-Love Number'':
\begin{equation}
    \text{SL}^{(LC)} = \frac{O(\nu T_{\alpha}h_{\beta} b^{\alpha\beta})}{O(\text{Bending Forces})} = \nu\text{.}
\end{equation}
As such, the dimensionless number associated with out of plane liquid crystalline dissipative forces is just the spin connection coefficient, $\nu$.

In addition to this, there is also the dimensionless active stress. As this is a number comparing tangential stress to bending stress, we can consider it as an analogue of the F\"oppl-von-K\'arm\'an number, $\text{Y}$, that would typically compare stretching to bending forces in passive shell theory \citep{Audoly2010}. In our case, this is given by
\begin{equation}
    \text{Y}^{(\text{Active})} = \frac{\Sigma L^2}{\mathcal{K}}\text{.}
\end{equation}
Note that this implies that in highly curved geometries bending forces will suppress active deformations.

\section{Applications}
To illustrate the potential interplay between geometry, ordering and flow, we provide some concrete examples. We focus first on the case where activity is not present, but where the highly curved geometry of a tube leads to couplings between the director and shape that are not present in the flat or close-to-flat cases studied in \citep{Salbreux22}. We then consider the ways in which active anisotropic stresses can drive tubes unstable, before finally considering the effects of activity on a topological defect located at the centre of a close to flat sheet.

\subsection{Relaxation dynamics about a perturbed tube}
We start with a simple example of a tube of radius $r(\theta,z,t)= r + \epsilon u(\theta,z,t)$, where $\epsilon$ is some dimensionless parameter that will be considered small. Such tubes have been shown in the case of isotropic membranes to have non-negligible Scriven-Love effects \citep{Sahu20}. The surface is prescribed by the vector $\bs{R}=(r\cos\theta,r\sin\theta,z)$. We can thus calculate the induced basis and metric
\begin{equation}
[g_{\alpha\beta}]=\left(\begin{matrix}
r(r+2\epsilon u) & 0\\
0 & 1
\end{matrix}\right)+O(\epsilon^2)\text{,}
\end{equation} 
and the second fundamental form
\begin{equation}
[b_{\alpha\beta}]=\left(\begin{matrix}
\epsilon  u_{,\theta\theta}-\epsilon  u-r & \epsilon u_{,\theta z }\\
\epsilon u_{,\theta z} & \epsilon u_{,zz}
\end{matrix}\right)+O(\epsilon^2)\text{.}
\end{equation}
This gives mean and Gaussian curvature to first order as
\begin{equation}
H=\frac{\epsilon}{2}\left( u_{,zz}+\frac{1}{r^2} u_{,\theta\theta}+  \frac{1}{r^2}u\right)-\frac{1}{2 r}\text{,}
\end{equation}
\begin{equation}
K=-\frac{\epsilon u_{,zz}}{r}\text{.}
\end{equation}

We write the director field as $\bs{T}=\frac{\epsilon T^\theta}{r}\bs{e}_\theta + (1+\epsilon T^z) \bs{e}_z +O(\epsilon^2)$, where we have chosen the $O(1)$ term to be the unit vector in the $z$ direction such that the molecular field vanishes for the unperturbed state \citep{Napoli12}. Thus, at lowest order, the constraint equation on director unity becomes
\begin{equation}\label{eq:unitNormLinearTube}
\partial_t T^z=0\text{,}
\end{equation}
and all the dynamics are in the $\theta$ direction, \textit{i.e.}~perpendicular to the initial zeroth order nematic ordering, as one would expect for the perturbative dynamics of a unit vector field.
The velocity we write as $\vec{V} =\epsilon (v^\theta/r\bs{e}_\theta +v^z\bs{e}_z + \partial_t u \bs{\hat{n}})$ and so the continuity equation becomes
\begin{equation}\label{eq:continuityLinearTube}
\frac{\partial_t u}{r}+\frac{v^\theta_{,\theta}}{r}+v^z_{,z}=0\text{.}
\end{equation}
The molecular field is given by
\begin{align}
h^\alpha\bs{e}_\alpha = &\mathcal{K}  \epsilon  \left(T^\theta_{,zz} + \frac{1}{r^2}T^\theta_{,\theta\theta} - \frac{1}{r^2}T^\theta + \frac{2}{r^2} u_{,\theta z}\right)\frac{1}{r}\bs{e}_\theta + \mathcal{K}   \epsilon  \left(\frac{T^z_{,\theta\theta}}{r^2} + T^z_{,zz}\right)\bs{e}_z\text{,}
\end{align}
and can be substituted to then find the polarization rate equation whose components are given by
\begin{align}\label{eq:textureLinearTube}
&\bs{e}_\theta:\quad \epsilon  \Bigg[- \frac{\mathcal{K} }{\gamma r}  \left(T^\theta_{,zz} + \frac{T^\theta_{,\theta\theta}}{r^2} - \frac{T^\theta}{r^2} + 2 \frac{u_{,\theta z}}{r^2}\right)+\frac{1}{r} \left( \partial_t T^\theta + \frac{1}{2} v^\theta_{,z} +\frac{1}{2r} v^z_{,\theta}\right)\nonumber\\
&\quad\quad\quad+ \frac{1}{r} \left( \lambda  T^\theta + \frac{\nu}{2}  v^\theta_{,z} + \frac{\nu}{2r}  v^z_{,\theta}\right)\Bigg]=0\text{,}\\
&\bs{e}_z:\quad \lambda-\epsilon\left[\frac{\mathcal{K} }{\gamma}  \left(\frac{1}{r^2}T^z_{,\theta\theta}+ T^z_{,zz}\right) +  \partial_t T^z + \lambda  T^z + \nu v^z_{,z}\right]=0\text{.}
\end{align}
The modified covariant Stokes equation (tangential force balance) is given component-wise by
\begin{align}\label{eq:stokesLinearTube}
&\bs{e}_\theta:\quad \frac{\zeta_{,\theta} (r-2 \epsilon  u)}{r^3}+\frac{\mathcal{K}  (\nu -1) \epsilon}{2 r^3}  \left(r^2 T^\theta_{,zzz} - T^\theta_{,z} + T^\theta_{,\theta\theta z} + 2 u_{,\theta zz}\right)\nonumber\\
&\quad\quad\quad+\frac{\eta  \epsilon}{r^3}  \left(r^2 v^\theta_{,zz} + \partial_t u_{,\theta} + v^\theta_{,\theta\theta}\right)=0\text{,}\\
&\bs{e}_z:\quad\frac{\eta  \epsilon}{r^2}  \left(r^2 v^z_{,zz}-r \partial_t u_{,z} + v^z_{,\theta \theta}\right)+\zeta_{,z} = 0\text{.}
\end{align}

The normal components of force due to the liquid crystal free energy are more complicated to calculate, however. Nevertheless, at $O(\epsilon)$ only the third term of (\ref{eq:normalLCforce}) contributes, giving
\begin{equation}
\left(b^\alpha{}_\gamma T^\gamma T^\beta\right)_{;\beta\alpha} = \epsilon\left[u_{,zzzz}-\frac{1}{r^2}T^\theta_{,z\theta} +\frac{1}{r^2}u_{,zz\theta\theta}\right]+O(\epsilon^2)\text{,}
\end{equation}
which gives the following shape equation
\begin{align}\label{eq:shapeLinearTube}
&\frac{\mathcal{K}  \epsilon}{r^2}  \left(r^2 u_{,zzzz} - T^\theta_{z\theta} + u_{,zz\theta\theta}\right) -\frac{\zeta}{r^2} \left[ \epsilon\left(r^2  u_{,zz} + u_{,\theta\theta} + u\right)-r\right]\nonumber\\
&\quad\quad+\frac{2 \eta  \epsilon}{r^2}  \left(\partial_t u + v^\theta_{,\theta}\right)=0\text{.}
\end{align}
Taking the case where $\epsilon=0$ we find a ground state with
\begin{align}
&\lambda=0\text{,}\\
&\zeta = 0\text{,}\\ 
&\bs{T}= \bs{e}_{z}\text{,}
\end{align}
as required. Note that as the bending energy term is anisotropic it gives no characteristic lengthscale of the tube as compared to the case of a lipid bilayer, for example, where the groundstate radius of the tube is set by the ratio of isotropic bending energy to surface tension ($r_0=\sqrt{\kappa/2\zeta}$) \citep{Zhong-Can89,Nelson95,Gurin96,Powers10,Boedec14,Fournier07}. 

We now take $\zeta=\epsilon\zeta$, $\lambda=\epsilon\lambda$ and, take the Fourier transform in space $\bar{f}_{q,m}=(1/4\pi^2)\int\mathrm{d}z\mathrm{d}\theta f(z,\theta)e^{-\dot\imath m \theta -\dot\imath q z}$. Using each of (\ref{eq:unitNormLinearTube}), (\ref{eq:continuityLinearTube}), (\ref{eq:stokesLinearTube}), (\ref{eq:textureLinearTube}) \& (\ref{eq:shapeLinearTube}), leaves simply two dynamical equations for $\bar{u}_{qm}$ and $\bar{T}^\theta_{qm}$ which are given in dimensionless form as follows
\begin{align}\label{eq:fullTubePassiveRelaxation}
\partial_{\tilde{t}} \left(\begin{matrix}
\bar{T}^\theta_{qm}\\
\bar{u}_{qm}
\end{matrix}\right)=A \cdot\left(\begin{matrix}
\bar{T}^\theta_{qm}\\
\bar{u}_{qm}
\end{matrix}\right) = \left(\begin{matrix} A_{11} & A_{12}\\
A_{21} & A_{22}\end{matrix}\right)\cdot\left(\begin{matrix}
\bar{T}^\theta_{qm}\\
\bar{u}_{qm}
\end{matrix}\right)\text{,}
\end{align}
where component-wise
\begin{align}
    &A_{11} = \frac{m^2 \left(-\left(Q^2 \left(4 \bar\eta +\nu ^2+1\right)+1\right)\right)-Q^2 \left(Q^2+1\right) \left(4 \bar\eta +(\nu -1)^2\right)}{4 Q^2}\text{,}\nonumber\\
    &A_{12} = \frac{m \left(m^4-2 m^2 \left((\nu -1) Q^2+1\right)+(1-2 \nu ) Q^4-2 Q^2 \left(4 \eta +(\nu -1)^2\right)\right)}{4 Q}\text{,}\nonumber\\
    &A_{21} = \frac{m \left(-2 \nu  Q^2 \left(m^2+Q^2+1\right)+m^2+Q^2\right)}{4 Q^3}\text{,}\nonumber\\
    &A_{22}= -\frac{m^6+m^4 \left(3 Q^2-2\right)+m^2 Q^2 \left(4 \nu +3 Q^2-2\right)+Q^6}{4 Q^2}\text{,}
\end{align}
where time has been normalized by $t=\tau\tilde{t}$ with $\tau=\eta r^2/\mathcal{K}$, $\bar{\eta}=\eta/\gamma$, $Q=rq$ and $\bar{u}$ has been non-dimensionalised by $r$. Note that the off-diagonal elements couple linearly in $mQ$. That is, the sign of $\bar{T}^{\theta}_{qm}$ selects a handedness of coupling to the shape deformations. 

\begin{figure}
\includegraphics[width=\textwidth]{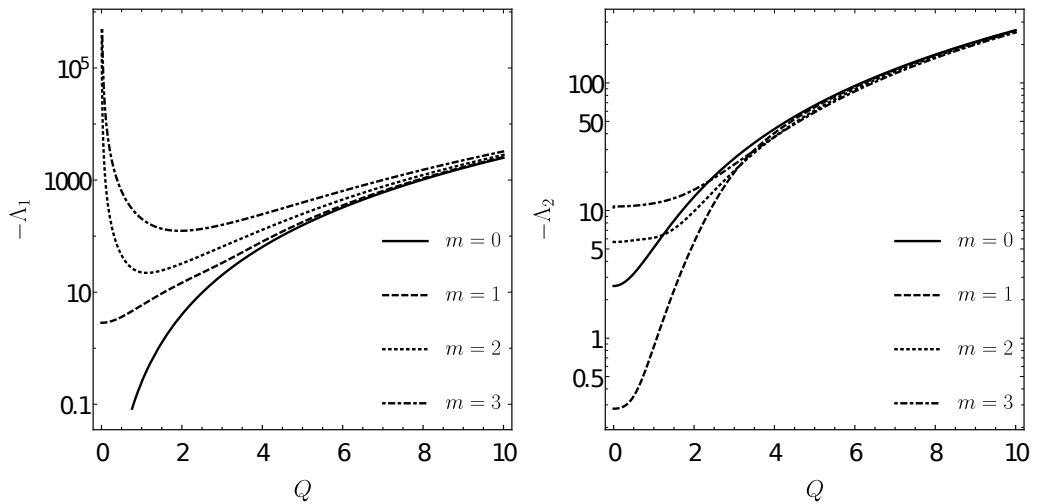}
\caption{\label{fig:passiveRelaxation}Relaxation rates for a perturbed liquid-crystal tube. $\Lambda_i$ are the two Eigenvalues of the growth matrix given in (\ref{eq:fullTubePassiveRelaxation}), that is, the growth rate of the spatial Fourier modes with time nondimensionalised by $\tau=\eta r^2/\mathcal{K}$ where $\eta$ is the $2$D viscosity, $r$ the tube radius and $\mathcal{K}$ the liquid crystalline splay modulus. These are plotted against dimensionless wavenumber $Q=q r$ ($z$ direction Fourier variable) for the first $4$ integer modes in $m$ (the $\theta$ Fourier variable). The dimensionless viscosity $\bar\eta=1$ and spin connection $\nu=-1.5$.}
\end{figure}

In contrast to the case of a flat membrane, this shows that in highly curved environments, the relaxation of shape and orientational ordering cannot be decoupled even in the case when the ground-state has no intrinsic curvature. This is true for all the modes on the tube with the exception of the $m=0$ case where the ordering and shape decouple. We plot the negative of the Eigenvalues of the matrix $A$, $\{\Lambda_1,\Lambda_2\}$ as a function of wavenumber $Q$ for the first $m=0,1,2,3$ modes in Fig.~\ref{fig:passiveRelaxation}. These can be interpreted as the passive relaxation modes that return perturbations back to the ground-state.

\subsection{Active Nematic Liquid Crystals}\label{sec:activeLCs}
 In this section, we will consider the effects of a non-zero active stress on the stability of nematic tubes, and on the morphology of a surface with a $+1$ defect located at the origin. In particular, we will examine the effects that changing the sign of the active stress (\textit{i.e.}~contractile vs.~extensile) on the morphology and surface instabilities.

\subsubsection{Active ruffling and pearling instabilities in tubes}
Using our equation for the active stress we can find normal part of the active force for a tube is given by
\begin{equation}
\bs{f}^a\cdot\bs{\hat{n}} = \sigma^{\text{(a)}\alpha\beta}b_{\alpha\beta}= \frac{\Sigma  \left[\epsilon\left(r^2  u_{,zz}-u_{,\theta\theta} -  u\right)+r\right]}{2 r^2}\text{,}
\end{equation}
and tangential parts by
\begin{equation}
\bs{f}^a\cdot\bs{e_\alpha}=\sigma^{\text{(a)}\alpha\beta}{}_{;\beta}\bs{e}_\alpha = \frac{\Sigma  \epsilon T^\theta_{,z}}{r}\bs{e}_\theta + \frac{\Sigma  \epsilon  \left(2 r T^z_{,z} + T^\theta_{,\theta} +u_{,z}\right)}{r}\bs{e}_z\text{.}
\end{equation}
Following a similar procedure as before we now find a different ground-state surface tension at $\epsilon=0$ for the unperturbed tube of $\zeta= \Sigma/2 +O(\epsilon)$. Now following the same procedure as before of Fourier transforming the equations around this ground-state we can find the dynamical equations for shape and alignment. Here we will consider only the axisymmetric deformations ($m=0$) for simplicity. This leads to the following equations for $\bar{u}_{q}$ and $\bar{T}^\theta_{q}$
\begin{align}
&\partial_{\tilde{t}}\bar{T}^{\theta}_{q}= -\frac{\left(\left(Q^2+1\right) \left((\nu -1)^2+4 \bar\eta \right)-2 (\nu -1) \bar\Sigma \right)}{4}\bar{T}^\theta_q\text{,}\\
&\partial_{\tilde{t}}\bar{u}_{q} = -\frac{\left( Q^4+\left(Q^2-1\right) \bar{\Sigma} \right)}{4}\bar{u}_q\text{,}
\end{align}
where $t=\tau\tilde{t}$ with $\tau=\eta r^2/\mathcal{K}$, $\bar{\eta}=\eta/\gamma$, $\bar{\Sigma}=\Sigma r^2/\mathcal{K}$, $Q=rq$ and $\bar{u}$ has been non-dimensionalised by $r$. Note that $\bar\Sigma$ is essentially a ratio of surface forces to bending energy and can thus be viewed as an ``Active F\"oppl-von-K\'arm\'an number'' that we discussed earlier.

Firstly we will consider the second equation describing the shape dynamics. The growth rate for the shape perturbation, $\bar{u}_q$, is given by
\begin{equation}
\Lambda_u(Q,\bar{\Sigma})=-\frac{\left( Q^4+\left(Q^2-1\right) \bar{\Sigma} \right)}{4}\text{.}
\end{equation}
This can be driven unstable for
\begin{equation}
\bar\Sigma\leq \frac{-Q^4}{Q^2-1}\text{.}
\end{equation}

\begin{figure}
\center\includegraphics[width=0.5\textwidth]{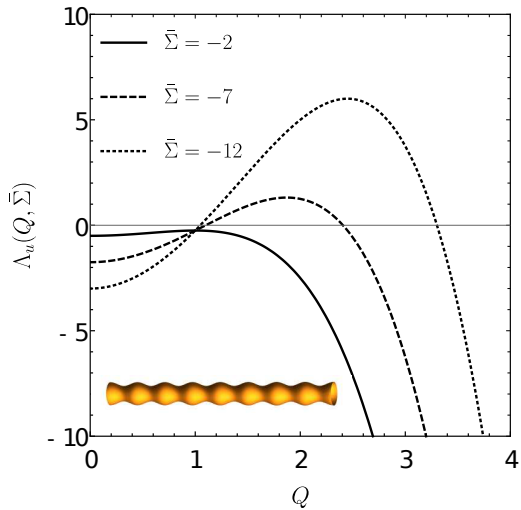}
\caption{\label{fig:extensileInstability}Extensile instability causes a ruffling-like deformation in the shape of the tube. $\Lambda_u(Q,\bar{\Sigma})$ is the growth rate of the spatial Fourier modes with time non-dimensionalised by $\tau=\eta r^2/\mathcal{K}$ where $\eta$ is the $2$D viscosity, $r$ the tube radius and $\mathcal{K}$ the liquid crystalline splay modulus. The active stress is non-dimensionalised as $\bar\Sigma=\Sigma r^2/\mathcal{K}$. Inset shows a representative example of such an instability.}
\end{figure}

For the case of an extensile active stress ($\Sigma<0$) this leads to an instability of a finite range of $Q>1$ peaked around a specific wavelength, see Fig.~\ref{fig:extensileInstability}. The fastest growing mode of this instability is found at $Q^\star=\sqrt{-\bar\Sigma/2}$ and as this at $Q>1$ leads to a ruffling-like undulations in the membrane. This makes sense heuristically as the extensile stresses are pointing along the tube thus generating forces which act to increase the surface area of the tube, hence the ruffling.

\begin{figure}
\includegraphics[width=\textwidth]{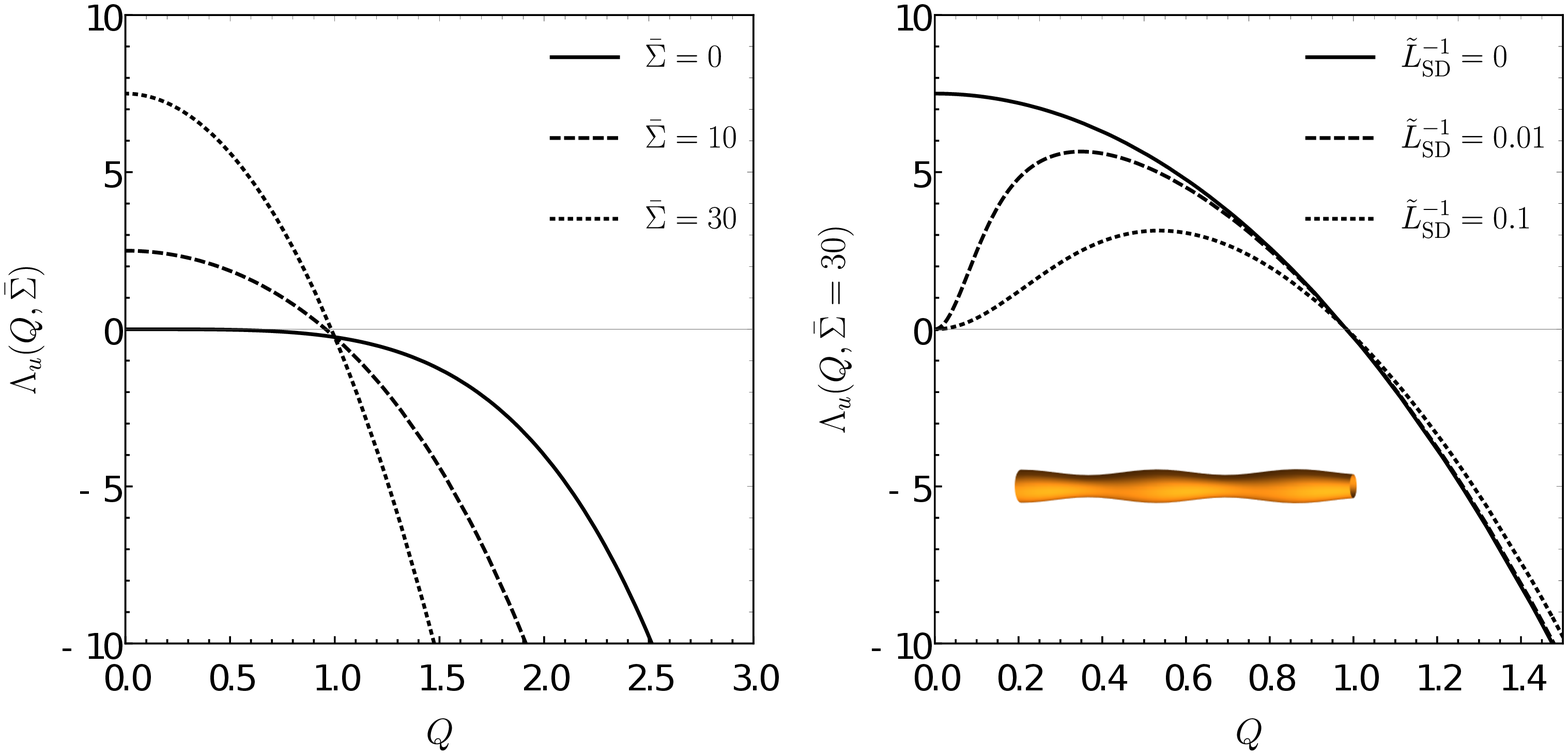}
\caption{\label{fig:contractileInstability}Contractile instability causes a pearling-like deformation in the shape of the tube. $\Lambda_u(Q,\bar{\Sigma})$ is the growth rate of the spatial Fourier modes with time non-dimensionalised by $\tau=\eta r^2/\mathcal{K}$ where $\eta$ is the $2$D viscosity, $r$ the tube radius and $\mathcal{K}$ the liquid crystalline splay modulus. The active stress is non-dimensionalised as $\bar\Sigma=\Sigma r^2/\mathcal{K}$. \textbf{Left:} instability in the absence of an ambient fluid. \textbf{Right:} instability when an approximation to ambient hydrodynamics in the long wavelength limit is used. $\tilde{L}_{\text{SD}}= \frac{\eta}{\eta_b r}$ is the dimensionless Saffaman-Delbr\"uck length given by ratio of $2$D to $3$D viscosity ($\eta_b$) non-dimensionalised by tube radius. Inset shows an example growth of a surface undulation for the case with bulk fluid.}
\end{figure}

In the case of contractile stresses ($\Sigma>0$) we find an instability with fastest growing mode in the $Q\to 0$ limit, Fig.~\ref{fig:contractileInstability} left. This instability is analogous to a classical Rayleigh-Plateau instability \citep{Rayleigh92,Tomotika35} and similar to the pearling instabilities seen in fluid membrane tubes \citep{Gurin96,Boedec14,Narsimhan15}. In reality such an instability does not occur at infinite length scales as the dissipative timescale in the ambient media damps the longer wavelength modes. We can approximate the bulk dissipation by modifying our shape dynamics equation in the following manner \citep{Gurin96,Boedec14,Powers10,Nelson95,Al-Izzi20b}
\begin{equation}
\Lambda_u(Q,\bar{\Sigma})=-\frac{\left( Q^4+\left(Q^2-1\right) \bar{\Sigma} \right)}{4+8 Q^{-2}\tilde{L}_{\text{SD}}^{-1}}\text{,}
\end{equation}
where $\tilde{L}_{\text{SD}} = \eta/(\eta_b r)$ is the dimensionless Saffman-Delbr\"uck length, and $\eta_b$ is the ambient viscosity \citep{Saffman75,Saffman76}. Here, this leads to a long wavelength instability as the contractile stress tries to minimise the surface to volume ratio of the tube. This gives a finite wavelength to the fastest growing mode set by the Saffman-Delbr\"uck length, Fig.~\ref{fig:contractileInstability} right. This instability is similar to the isotropic contractile instabilities previously found in fluid membranes \citep{Mietke19}.

\begin{figure}
\center\includegraphics[width=\textwidth]{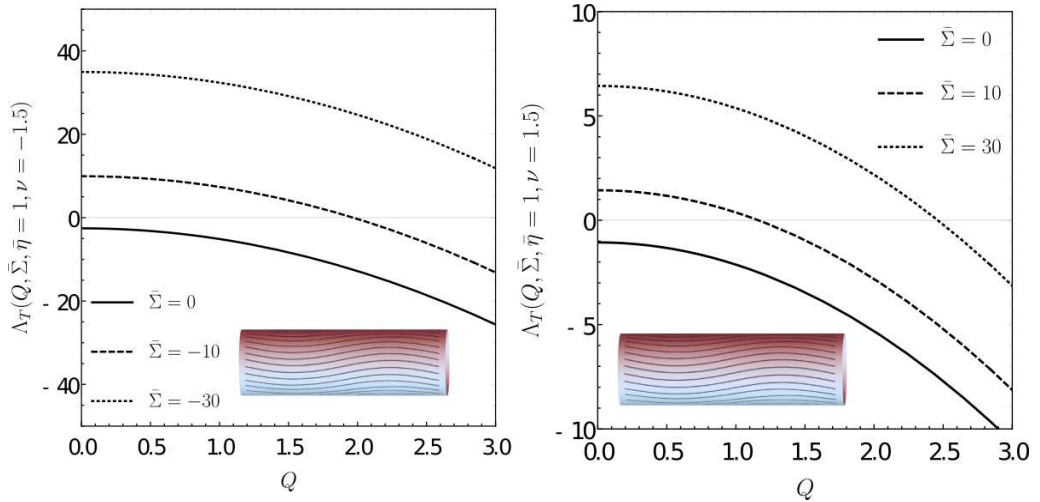}
\caption{\label{fig:bendInstability}\textbf{Left:} Extensile bend instability in texture on the surface of the tube with rod-like nematics. \textbf{Right:} Contractile bend instability in texture on the surface of the tube with disk-like nematics. $\Lambda_T(Q,\bar{\Sigma})$ is the growth rate of the spacial Fourier modes with time non-dimensionalised by $\tau=\eta r^2/\mathcal{K}$ where $\eta$ is the $2$D viscosity, $r$ the tube radius and $\mathcal{K}$ the liquid crystalline splay modulus. The active stress is non-dimensionalised as $\bar\Sigma=\Sigma r^2/\mathcal{K}$. The dimensionless viscosity $\bar\eta=1$ and spin connection $\nu=-1.5$ (Left, rod-like nematics) and $\nu=1.5$ (Right, disk-like nematics). Insets shows an example of the bend instability at the longest wavelength of the tube.}
\end{figure}

We now turn our attention to the growth rate of the director deviations, which is given by
\begin{align}
\Lambda_T(Q,\bar{\Sigma},\bar{\eta},\nu)=-\frac{\left(\left(Q^2+1\right) \left(4 \bar\eta +(\nu -1)^2\right)-2 (\nu -1) \bar\Sigma \right)}{4}\text{.}
\end{align}
This can also have a change in sign when in both the extensile or contractile regime, depending on the sign of $\nu$ ($\nu<-1$ corresponds to rod-like nematics and $\nu>1$ to disk-like). The stability condition is given by
\begin{equation}
\bar\Sigma\geq \frac{(Q^2+1)\left(4\bar\eta+(\nu-1)^2\right)}{2(\nu-1)}\text{,}
\end{equation}
which leads to a spontaneous bend instability in the longest wavelengths of the tube, Fig.~\ref{fig:bendInstability}. This instability is well known in flat geometry in the case of extensile stress and rod-like nematics ($\nu<-1$) where it is known to lead to active turbulence \citep{Marchetti13}.

In the limit $Q\to 0$, this criterion becomes $\bar\Sigma \geq \left(4\bar\eta+(\nu-1)^2\right)/2(\nu-1)$ in comparison to the flat case where the instability is threshold-less \cite{Marchetti13}. This is directly due to the correct use of the surface derivative as the texture needs sufficient active stress to overcome the elastic stress the bend induces by forcing the texture to bend around the tube. Because of this threshold it is interesting to note that the shape instability precedes the bend instability, suggesting it should be possible to generate shape changes before one sees active turbulence. The interplay between this instability and the shape equations beyond linear order is likely a very rich topic as this director instability would likely couple to helical deformation modes leading to a spontaneous chiral symmetry breaking. However such a topic is beyond the scope of the analysis in this paper.

\subsubsection{Active topological defect deformations}
Here we consider a planar membrane in polar Monge coordinates, where the surface is parameterised by the Cartesian vector $\bs{R}=\left(r\cos\theta,r\sin\theta,\epsilon z\right)$. For simplicity we will only consider axisymmetric steady-states, thus $z(r,\theta)=z(r)$. The metric and second fundamental form are given as
\begin{equation}
[g_{\alpha\beta}] = \left(
\begin{array}{cc}
 1 & 0 \\
 0 & r^2 \\
\end{array}
\right) +O(\epsilon^2)\text{,}
\end{equation}
\begin{equation}
[b_{\alpha\beta}] = \left(
\begin{array}{cc}
 \epsilon  z_{,rr} & 0 \\
 0 & r \epsilon  z_{,\theta} \\
\end{array}
\right) + O(\epsilon^2)\text{,}
\end{equation}
up to first order in $\epsilon$. Thus we find $H= \frac{\epsilon}{2}\left(\frac{1}{r}z_{,r}+z_{,rr}\right)+O(\epsilon^2)$ and $K=0+O(\epsilon^2)$. We take the polarisation texture to be $\bs{T}=\cos\phi\bs{e}_{r} + \frac{1}{r}\sin\phi\bs{e}_\theta$ where $\phi$ is a constant such that there is a $+1$ topological defect located at $r=0$. The continuity equation then yields $v^r=0$, so we are left to solve for $z(r,t)$, $\zeta(r,t)$, $\lambda(r,t)$, $v^\theta(r,t)$.

At lowest order in $\epsilon$ the the equations are independent of $\dot{z}$ so we look for steady states in the nematic order parameter. Our choice of $\bs{T}$ trivially satisfies the unit magnitude constraint. We now make use of the fact that, at lowest order the tangential force balance and polarization equations are just given by their flat counterparts (that is, they have no $O(\epsilon)$ term) and the shape equation is given entirely by $O(\epsilon)$ terms \citep{Kruse05,Hoffmann21}. 

The polarization equations become
\begin{align}
&(\nu  \cos (2 \phi )-1) \left( v^\theta_{,r}-\frac{v^\theta}{r}\right)=0\text{,}\\
&\nu  \sin (\phi ) \cos (\phi ) \left( v^\theta_{,r}-\frac{v^\theta}{r}\right)+\lambda +\frac{\mathcal{K} }{\gamma r^2}=0\text{,}
\end{align}
which, assuming $\nu>1$ (disk-like nematics), gives
\begin{align}
&\phi = \pm \frac{1}{2}\arccos\left[\frac{1}{\nu}\right]\text{,}\\
& \lambda =- \frac{\mathcal{K}}{r^2 \gamma} - \nu\sin (\phi)\cos(\phi)\left[v^\theta_{,r}-\frac{v^\theta}{r}\right]\text{.}
\end{align}
Substituting the solution for the $\phi$ into the tangential force balance gives
\begin{align}
&\zeta_{,r} + \frac{\mathcal{K}  \nu }{r^3}+\frac{\Sigma }{\nu  r} = 0\text{,}\\
&r \left(\eta  v^\theta_{,r} + \eta  r v^\theta_{,rr} \pm \sqrt{1-\frac{1}{\nu ^2}} \Sigma \right)-\eta  v^\theta = 0\text{,}
\end{align}
for which solutions are
\begin{align}
&\zeta = \frac{\mathcal{K}\nu}{2 r^2} - \frac{\Sigma}{\nu} \log\left[\frac{r}{r_{\text{max}}}\right]+\zeta_0\text{,}\\
&v^\theta = \mp \frac{\Sigma}{2\eta} \sqrt{1-\frac{1}{\nu^2}} r \log\left[\frac{r}{r_{\text{max}}}\right]\text{,}
\end{align}
where $\zeta_0 + \mathcal{K}\nu/2 r^2_{\text{max}}$ is the surface tension at $r=r_{\text{max}}$. This solution is identical to that found in \cite{Kruse04} and is one of the classical results of polar active gel theory in a flat geometry. Notice that the surface tension diverges as $r\to 0$ due to the presence of the defect. This divergence in shape can be avoided by introducing a short wavelength cut-off, $r_c$, or by considering a more general Landau-de-Gennes free energy that allows for a nematic isotropic phase transition \citep{DeGennes93}.

\begin{figure}
\center\includegraphics[width=\textwidth]{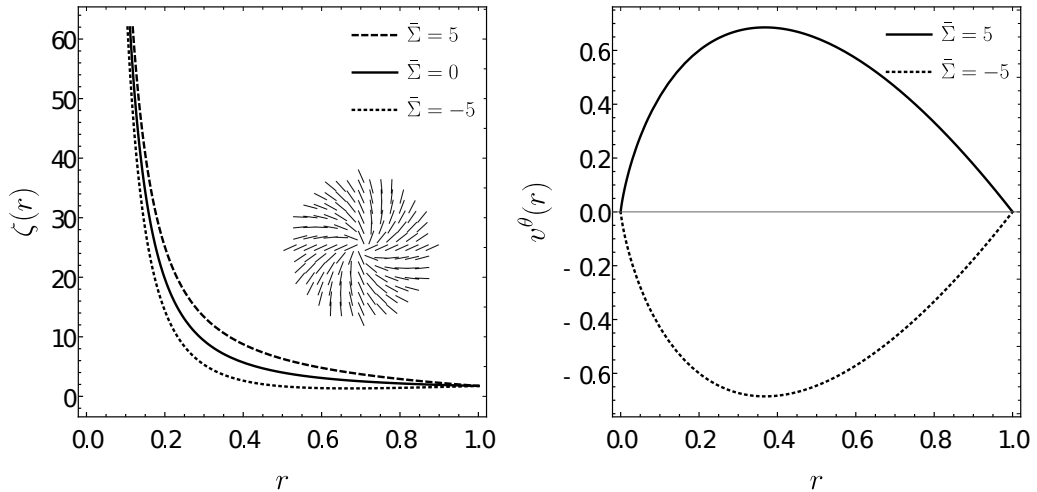}
\caption{\label{fig:defectFlows}\textbf{Left:} Defect surface tension, $\zeta(r)$, as a function of radius, $r$, about a $+1$ defect centered on $r=0$. Inset is defect texture. \textbf{Right:} Velocity field around the defect. Here we have introduced a small cut-off radius at $r=0.03$ where lengths are non-dimensionalised by the radius of the boundary $r_{\text{max}}=1$. Surface tension is given by $\bar\zeta_0=\zeta_0 r_{\text{max}}^2/\mathcal{K}=1.0$, spin connection $\nu=1.5$, $r_c=0.03$ and $c=-0.05$ where $c$ essentially defines the contact angle at the boundary. The active stress is non-dimensionalised by $\bar\Sigma = \sigma r_{\text{max}}/\mathcal{K}$. We plot for both positive (contractile) and negative (extensile) values of active stress to show promotion/suppression of the deformation in each case.}
\end{figure}

Substituting our solution for the surface tension into the normal force balance equation yields the following dimensionless shape equation
\begin{align}
    &\frac{z_{,r}}{r^3} \left(7-\nu  \left(2 \bar\zeta_0 r^2+5\right)+2 r^2 \bar\Sigma  \log (r)+r \left(\sqrt{\nu ^2-1}+r \bar\Sigma \right)\right)\nonumber\\
    &-\frac{z_{,rr}}{r^2} \left(r^2 (2 \bar\zeta_0 \nu + \bar\Sigma )-2 r^2 \bar\Sigma  \log (r)+4\right)\nonumber\\
    &+ (\nu +1) \left( z_{,rrrr} + \frac{2}{r} z_{,rrr}\right)=0\text{,}
\end{align}
where lengths have been non-dimensionalised by $r_{\text{max}}$ and energies by $\mathcal{K}$. We can solve this equation numerically with from the boundary ($r=1$ in our dimensionless units) up to a small cut-off, $r_c$. We solve with zero mean curvature and derivative in curvature at $r=r_{\text{max}}=1$, at lowest order this is $z'(r_{\text{max}})=-z''(r_{\text{max}})=z'''(r_{\text{max}})=c=\text{Const.}$ and fix $z(r_{\text{max}})=0$ at the boundary. The constant, $c$, essentially prescribes the contact angle at the boundary. To solve this numerically we make use of Mathematica (WolframResearch, Champaign, IL).

\begin{figure}
\center\includegraphics[width=\textwidth]{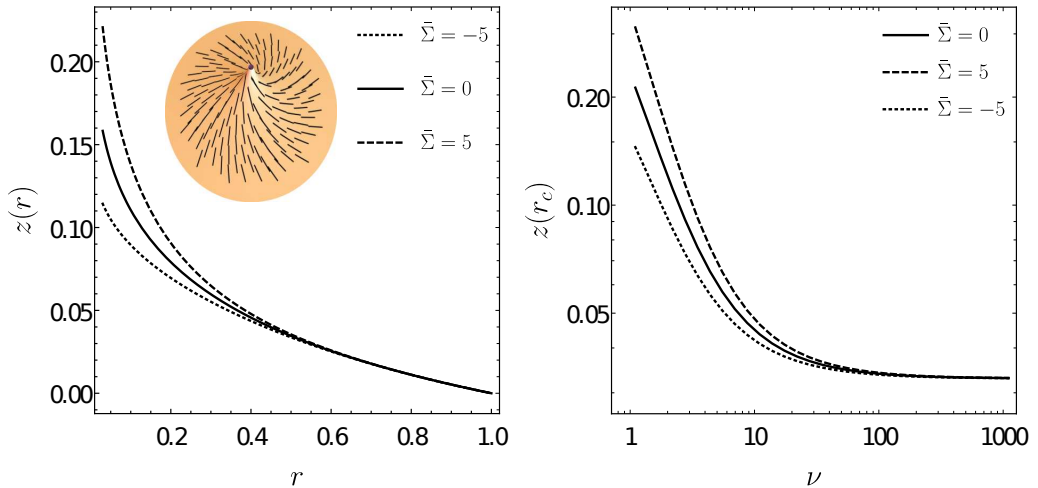}
\caption{\label{fig:defectHeight}\textbf{Left:} Surface heights, $z(r)$, as a function of radius, $r$, about a $+1$ defect centered on $r=0$. Here we have introduced a small cut-off radius at $r=0.03$ where lengths are non-dimensionalised by the radius of the boundary $r_{\text{max}}=1$. Surface tension is given by $\bar\zeta_0=\zeta_0 r_{\text{max}}^2/\mathcal{K}=1.0$, spin connection $\nu=1.5$, $r_c=0.03$ and $c=-0.05$ where $c$ essentially defines the contact angle at the boundary. The active stress is non-dimensionalised by $\bar\Sigma = \sigma r_{\text{max}}/\mathcal{K}$. We plot for both positive (contractile) and negative (extensile) values of active stress to show promotion/suppression of the deformation in each case. Inset is an example surface with the texture shown on top. \textbf{Right:} Maximum defect height, $z(r_c=0.03)$, plotted against spin connection $\nu$ (or liquid crystalline Scriven-Love number) in the passive, contractile and extensile cases.}
\end{figure}

Solutions to this for varying values of active stress are plotted in Fig.~\ref{fig:defectHeight} with $\bar\zeta_0=1.0$, $\nu=1.5$, $r_c=0.03$ and $c=-0.05$. For $\bar\Sigma=0$ we find a convex cusp solution where the stress from the defect causes the surface to buckle. Note that this is qualitatively similar to the full non-linear pseudosphere cusp found in \citep{Frank08} although the quantitative details of our shape differ significantly due to our inclusion of the extrinsic curvature term in the Frank free energy, and thus the anisotropic bending energies in our shape equation ($4{^\text{th}}$ and $3^{\text{rd}}$ order terms) making a full analytical non-linear solution of our equations impossible. For the extensile case, the protrusion decreases in height. The reason for this is directly due to our inclusion of the extrinsic terms in the Frank free energy and the convex shape of the passive solution. Extensile stresses lead to active forces which act to increase bend, in this case bend due to the deformation along the surface, thus pushing the surface down and suppressing the deformation. The opposite is true in the contractile case where the active forces act to flatten bend, thus pushing the surface upward. A simple example of these types effects was discussed in the context of an active fingering instability in \cite{Alert22}. In addition, we show that as the spin connection, $\nu$, (or liquid crystalline Scrinev-Love number) is increased and the spiral defect texture goes from aster-like to vortex-like as $\nu$ goes to $\infty$. As this happens the bend along the surface goes to zero and the height of the defect, $z(r_{c})$, converges in the extensile, contractile and passive cases, as expected. 

Note that our results here are different from those seen in \cite{Hoffmann21} where the effects of contractile and extensile forces are switched. This is straightforward to understand since \cite{Hoffmann21} did not consider the extrinsic coupling of the director and in addition included an isotropic bending energy. This causes the passive shape to be regularised around the defect and be concave  near the defect where the active forces are strongest, leading to a switch in sign of the normal force due to activity when compared to our convex passive shape. Such subtle interplay between passive and active stresses suggests a rich phenomonology of possible non-linear solutions to such systems, and that it may be possible for both extensile and contractile systems to achieve a similar morphology by the varying of passive parameters.

It is interesting to note that the contractile protrusions we find are similar to those seen during \textit{Hydra} morphogenesis in the underlying contractile actomyosin network \citep{Maroudas-Sacks21,Vafa21}. There, the $+1$ defects form the basis of protrusions which develop into ``limbs''.

\section{Discussion}
In this paper we have developed a fully covariant hydrodynamic theory of active nematic fluids on deformable surfaces, deriving equations for normal and tangential force balance along with an equation for order parameter dynamics. Focusing on the case of the one-constant Frank free energy, we identify three dimensionless numbers: two Scriven-Love numbers associated with the ratio of normal viscous forces to bending forces, and an active analogue of the F\"oppl-von-K\'arm\'an number, comparing tangential active stress to bending forces.

We then consider the relaxation dynamics, in shape and order parameter, of a nematic tube with no active forcing, showing that there is a non-trivial coupling between shape and director relaxation in the case of the non-axisymmetric modes. Motivated by the recent interest in active nematic fluids as models of morphogenetic processes, we further consider the effect an active $Q$-tensor term on tube morphology. Here we show that there are several new instabilities in both the contractile and extensile cases, which we compare to similar cases in flat geometry liquid crystals and isotropic fluid membranes. Finally we consider the effect of activity on the surface morphology around a $+1$ topological defect, where anisotropic stresses drive or suppress protrusion of the defect, dependent on whether stresses are contractile or extensile, respectively.

An interesting extension of the problems considered here would be the effect of $1/2$-integer defects on the shape. In addition to breaking axisymmetry, this presents an additional complication in that $+1/2$ integer defects are known to self-propel in active liquid crystals \citep{Giomi14}. We speculate that such effects would persist (at least in the small deformation limit) and would lead to self-propelled travelling waves in the shape. We aim to explore this phenomenology in detail in future work.

More generally, we believe the framework developed here has applications in a wide range of systems, bridging length and time scales. Our chief interest throughout was to develop the equations in a clear, systematic way that laid bare the basic phenomonology without recourse to biological specifics. With the addition of relevant concentration fields describing morphogens, growth factors and other signalling molecules, these equations could form the basis for a nematic theory of deformable epithelial tissues \citep{Al-Izzi21,Julicher18}. There are also additional areas of potential application in the field of lipid bilayer dynamics; when augmented with isotropic bending energy terms these equations could form the basis for a dynamic theory of tilt-chiral lipid bilayers, extending earlier work considering only energetics \citep{Selinger96}. It is also plausible that this is the natural framework to model a fluid bilayer coupled to a thin layer of active filaments \textit{e.g.}~actin \citep{Simon19}, microtubules \citep{Keber14} or engineered filaments such as DNA filaments \citep{Jahnke22}.

A major open challenge in the field of covariant hydrodynamics is developing general stable numerical methods to provide solutions beyond the linear perturbation regime or simple axisymmetric cases \citep{Mietke19}. A large body of work exists for such problems in the case of passive and active isotropic fluids based on unfitted finite elements, arbitrary Lagrangian-Eulerian finite elements or isogeometric analysis \citep{Barrett16,Torres-Sanchez19,Vasan20,Sahu20a}, but, to our knowledge, this has yet to be extended to nematic fluids on moving curved surfaces. It is important to note that the equations derived here would likely need to be modified to describe the $Q$-tensor dynamics rather than the nematic order parameter $\bs{T}$ so as to correctly deal with $1/2$ integer defects numerically. Solving such tensor PDEs on curved surfaces is challenging, with issues relating to the discontinuity of basis functions across elements. Although some progress has been made recently in approximating tensor fields numerically using local Monge approximations \citep{Torres-Sanchez20}.

Current methods for solving nemato-hydrodynamics are often hybrid lattice Boltzmann codes \citep{Denniston01,Binysh20} and such methods have been used along with phase fields to approximate interface dynamics \citep{Metselaar19,Hoffmann21}. We believe it would also be useful to develop tools to solve the full covariant equations of a deformable mesh using finite element methods so as to provide numerical methods which more clearly map to geometric analytical framework, and thus provide a clearer conceptual link between hydrodynamics and geometry. We therefore welcome further work in this area.

\section*{Declaration of interests}
The authors report no conflict of interest.

\section*{Acknowledgements}
The authors acknowledge support from the EMBL-Australia program and helpful comments and discussions with A.~Sahu (Cornell), G.P.~Alexander (Warwick), F.~Vafa (Harvard) \& J.~Binysh (Bath).

\appendix
\section{Preliminary differential geometry}\label{sec:diffGeoApp}
In this section we define some preliminary objects from differential geometry which we will make use of in the main derivation. We consider a surface, $\mathcal{M}_t\subset\mathbb{R}^3$ which is locally isomorphic to $\mathbb{R}^2$ and changes continuously in time (\textit{i.e.} the geometry of our interface throughout time is given by a one-parameter family of diffeomorphisms). This assumption neglects any topological changes in the surface and for simplicity we will also consider surfaces without boundary.

We define the embedding of $\mathcal{M}_t$ in $\mathbb{R}^3$ by the vector $\bs{R}(x_1,x_2)$ where we assume that locally one can write this as a function of two parameters $(x_1,x_2)$. Within this framework it is possible to define an induced tangent basis on the manifold by taking derivatives with respect to these coordinates $\bs{e}_\alpha=\bs{R}_{,\alpha}$ where the subscript, $(,)$, denotes partial differentiation. Note that in general these basis vectors are neither unit, nor orthogonal. From here we can define an induced metric as follows,
\begin{equation}\label{eq:metricDef}
g_{\alpha\beta} = \bs{R}_{,\alpha}\cdot\bs{R}_{,\beta} =\bs{e}_{\alpha}\cdot\bs{e}_{\beta}
\end{equation}
where $g_{\alpha\beta}$ is the metric tensor. It has an inverse, $g^{\alpha\beta}$, defined by $g^{\alpha\beta}g_{\alpha\gamma} = \delta^\beta{}_\gamma$, it defines an inner product with respect to two vectors in the tangent bundle of $\mathcal{M}_t$, and, as such defines a mapping between the tangent bundle and cotangent bundle on $\mathcal{M}_t$. This latter point means we can use the metric to map between vectors $\bs{x}=x^\alpha\bs{e}_\alpha$ and covectors $x = x_\alpha \bs{e}^\alpha = x^\beta g_{\alpha\beta}\bs{e}^\alpha$, \textit.{i.e.}~we can use the metric to raise and lower indices. 

The metric is sufficient to describe all intrinsic properties of a manifold, however if we also want to know about extrinsic properties, as infact we must if we want to describe how our surface is embedded within $\mathbb{R}^3$ then we also require knowledge about how the normal to the surfaces changes. We define a unit normal vector as follows $\bs{\hat{n}}=\bs{e}_1\times\bs{e}_2/|\bs{e}_1\times\bs{e}_2|$ where $\times$ is the cross produce in $\mathbb{R}^3$. The extrinsic curvature is then defined as the negative of the rate-of-change of the normal in the direction of the tangent basis, and using orthogonality of $\bs{\hat{n}}$ with $\bs{e}_\alpha$ we can write
\begin{equation}\label{eq:2ndFundamentalDef}
b_{\alpha\beta} = \bs{e}_{\alpha,\beta}\cdot\bs{\hat{n}}
\end{equation}
where mean and Gaussian curvatures are given by $H=b^\alpha{}_\alpha/2$ and $K=\mathrm{det}(b^\alpha{}_\beta)$ respectively.

Using orthogonalisty of the normal with the basis vectors and dotting in $\mathbb{R}^3$ with $\bs{e}_{\alpha}$ we find the Weingarten relation
\begin{equation}\label{eq:weigarten}
\bs{\hat{n}}_{,\alpha} = - b^{\beta}{}_{\alpha}\bs{e}_{\beta}
\end{equation}

And defining the connection $\Gamma^\gamma_{\beta\alpha} = \bs{e}_{\alpha,\beta}\cdot\bs{e}^{\gamma}$ we can find Gauss's formula
\begin{equation}\label{eq:gaussFormula}
\bs{e}_{\alpha,\beta} = \Gamma^{\gamma}_{\alpha\beta}\bs{e}_{\gamma} + b_{\alpha\beta}\bs{\hat{n}}
\end{equation}
and we thus define covariant differentiation in the tangent bundle in terms of this connection. The covariant derivative of a rank $(p,q)$ tensor is given by
\begin{align}
    \nabla_{\gamma}T^{\alpha_{1}\dots\alpha_p}{}_{\beta_1\dots\beta_q} =& T^{\alpha_{1}\dots\alpha_p}{}_{\beta_1\dots\beta_q,\gamma} + \Gamma^{\alpha_1}_{\gamma\delta} T^{\delta\dots\alpha_p}{}_{\beta_1\dots\beta_q} +\dots + \Gamma^{\alpha_p}_{\gamma\delta} T^{\alpha_{1}\dots\delta}{}_{\beta_1\dots\beta_q}\nonumber\\
    &- \Gamma^{\delta}_{\gamma\beta_1} T^{\alpha_{1}\dots\alpha_p}{}_{\delta\dots\beta_q}- \dots - \Gamma^{\delta}_{\gamma\beta_q} T^{\alpha_{1}\dots\alpha_p}{}_{\beta_1\dots\delta}\text{.}
\end{align}

On a torsion free Reimannian manifold the connection is simply given by the Christoffel symbols, which can be written in terms of the metric as
\begin{equation}
\Gamma^\gamma_{\alpha\beta} = \frac{1}{2} g^{\gamma\delta} \left( g_{\delta\beta,\alpha} +g_{\delta\alpha,\beta} - g_{\alpha\beta,\delta}\right)\text{.}
\end{equation}

A standard measure of intrinsic curvature of a manifold  is given by the Reimann tensor, which essentially measures the commutation of two covariant derivatives on a vector (and is thus intimatly related with parallel transport of vectors on surfaces). The Reimann tensor is defined as
\begin{equation}
R^\tau{}_{\alpha\gamma\beta} = \Gamma^\tau_{\alpha\beta,\gamma} - \Gamma^\tau_{\alpha\gamma,\beta} + \Gamma^\delta_{\alpha\beta}\Gamma^\tau_{\delta\gamma} - \Gamma^\delta_{\alpha\gamma}\Gamma^\tau_{\delta\beta}\text{.}
\end{equation}

For a $d$-dimensional surface equipped with a metric and extrinsic curvature tensor to be embeddable in a $d+1$ dimensional manifold the metric and second fundamental form must satisfy some constraints. These constrains are called the Gauss-Mainardi-Codazzi-Peterson equations, and in the case of a $2$D surface embedded in $\mathbb{R}^3$ are given by
\begin{align}
&R^\tau{}_{\alpha\beta\gamma} = b^\tau{}_{\gamma} b_{\alpha\beta} - b^\tau{}_\beta b_{\alpha\gamma} \label{eq:gauss}\\
& b_{\alpha\beta;\gamma}-b_{\alpha\gamma;\beta}=0\text{.}\label{eq:codazzi}
\end{align}

With some index manipulation one can show that in $2$D the Ricci tensor (a contraction of the Reimann tensor), and Ricci scalar are related to the Gaussian curvature by
\begin{align}
&R_{\alpha\beta} = R^\gamma{}_{\alpha\gamma\beta} = K g_{\alpha\beta}\\
&R=R_{\alpha\beta}g^{\alpha\beta} = 2 K\text{.}
\end{align}

Using this and contracting the Gauss equation (\ref{eq:gauss}) gives
\begin{equation}\label{eq:contractedGauss2D}
Kg_{\alpha\beta} = 2 H b_{\alpha\beta} - b_{\alpha}{}^{\gamma}b_{\gamma\beta}\text{.}
\end{equation}

The manifold $\mathcal{M}_t$ is under a flow given by the velocity field $\bs{V}=v^\alpha\bs{e}_{\alpha} + v^{(n)} \bs{\hat{n}}\in\mathbb{R}^3$. We can calculate the rate-of-change of the basis vectors under this flow as follows
\begin{align}\label{eq:coordinateRate}
\frac{\mathrm{d}}{\mathrm{d}t}\left(\bs{e}_{\alpha}\right)&=\bs{V}_{,\alpha} = v^\beta{}_{,\alpha}\bs{e}_{\beta} + v^\beta \bs{e}_{\beta,\alpha} + v^{(n)}{}_{,\alpha} \bs{\hat{n}} + v^{(n)} \bs{\hat{n}}_{,\alpha}\nonumber\\
 &= \left(v^\beta{}_{;\alpha} - v^{(n)} b^{\beta}{}_{\alpha}\right)\bs{e}_{\beta} + \left(v^\beta b_{\alpha\beta} + v^{(n)}{}_{;\alpha}\right)\bs{\hat{n}}
\end{align}
where we have used the Weingarten equation, $\bs{\hat{n}}_{,\alpha} = - b^{\beta}{}_{\alpha}\bs{e}_{\beta}$, and the Gauss formula, $\bs{e}_{\alpha,\beta} = \Gamma^{\gamma}_{\alpha\beta}\bs{e}_{\gamma} + b_{\alpha\beta}\bs{\hat{n}}$. Note that here we have included terms that are advected with the tangential flow, these are important when considering quantities defined in a Lagrangian frame (\textit{e.g.} the rate-of-deformation tensor), however, when considering objects defined in an objective/Eulerian frame we will only account for changes due to the normal velocity (as in changes that explicitly change the surface geometry). There are more concise ways to write these derivatives in a mixed Lagrangian-Eulerian manner, but this makes use of Lie derivatives so we avoid this here for the sake of simplicity.

We now proceed to calculate the time derivative of the normal vector, $\bs{\hat{n}}$, the first point to note is that $\frac{\mathrm{d}}{\mathrm{d}t}(|\bs{\hat{n}}|^2) =0 \Rightarrow \bs{\hat{n}}\cdot\frac{\mathrm{d}}{\mathrm{d}t}(\bs{\hat{n}}) = 0$ \textit{i.e.}~there are no normal components. Next using the fact that $\bs{e}_{\alpha}\cdot\bs{\hat{n}}=0$ (by construction),  we find the following 
\begin{align}\label{eq:normalRate}
\frac{\mathrm{d}}{\mathrm{d}t}\left(\bs{\hat{n}}\right) = - \bs{V}_{,\alpha}\cdot\bs{\hat{n}}g^{\alpha\beta}\bs{e}_{\beta} = - \left(v^\beta b_{\beta}{}^{\alpha} + v^{(n)}{}_{;\alpha} g^{\alpha\beta}\right)\bs{e}_{\alpha}\text{.}
\end{align}

In deriving the exact form of each of the functional derivatives the following geometric identities will prove useful. Throughout this paper we will make use of rate equations for geometric quantities in order to compute functional derivatives as we believe this simplifies computations somewhat as we will only be computing first variations in the energy.

Starting from the definition of the second fundamental form, (\ref{eq:2ndFundamentalDef}), we have
\begin{equation}
\frac{\mathrm{d}}{\mathrm{d}t}\left(b_{\alpha\beta}\right)= \left(\frac{\mathrm{d}}{\mathrm{d}t}\bs{e}_{\alpha,\beta}\right)\cdot\bs{\hat{n}} + \bs{e}_{\alpha,\beta}\cdot\frac{\mathrm{d}}{\mathrm{d}t}\bs{\hat{n}}\text{.}
\end{equation}

Examining the first term we find
\begin{equation}
\frac{\mathrm{d}}{\mathrm{d}t}\bs{e}_{\alpha,\beta}\cdot\bs{\hat{n}} = \bs{V}_{,\alpha\beta}\cdot\bs{\hat{n}}
\end{equation}
where
\begin{align}
\bs{V}_{,\alpha\beta} &= \Big[\left(v^\gamma{}_{;\alpha}-v^{(n)}b^\gamma{}_{\alpha}\right)\bs{e}_{\gamma} + \left(v^\gamma b_{\alpha\gamma} +v^{(n)}{}_{,\alpha}\right)\bs{\hat{n}}\Big]_{,\beta}\nonumber\\
& = \left(v^\gamma{}_{;\alpha}-v^{(n)}b^\gamma{}_{\alpha}\right)_{,\beta}\bs{e}_{\gamma} + \left(v^\gamma{}_{;\alpha}-v^{(n)}b^\gamma{}_{\alpha}\right)\bs{e}_{\gamma,\beta}\nonumber\\
& +\left(b_{\alpha\gamma}v^\gamma\right)_{,\beta} \bs{\hat{n}} + v^{(n)}{}_{,\alpha\beta}\bs{\hat{n}} +\left(v^\gamma b_{\alpha\gamma} + v^{(n)}{}_{,\alpha}\right)\bs{\hat{n}}_{,\beta}
\end{align}
Dotting with $\bs{\hat{n}}$ and making use of the Gauss's formula and Weingarten equation (\ref{eq:gaussFormula}) \& (\ref{eq:weigarten}) we find
\begin{equation}
\bs{V}_{,\alpha\beta}\cdot\bs{\hat{n}} = \left(v^\gamma{}_{;\alpha}-v^{(n)} b^\gamma{}_\alpha\right)b_{\gamma\beta} + \left(b_{\alpha\gamma}v^\gamma\right)_{,\beta} + v^{(n)}{}_{,\alpha\beta}\text{.}
\end{equation}

Making use of the equation for the normal vector dynamics, (\ref{eq:normalRate}), dotting with $\bs{e}_{\alpha,\beta}$, and making use of the Gauss's formula we find
\begin{equation}
\bs{e}_{\alpha,\beta}\cdot\frac{\mathrm{d}}{\mathrm{d}t}\bs{\hat{n}} = -\Gamma^\gamma_{\alpha\beta}v^\delta b_{\gamma\delta} - \Gamma^\gamma_{\alpha\beta}v^{(n)}_{,\gamma}\text{.}
\end{equation}

The equation for the time derivative of $b_{\alpha\beta}$ then becomes
\begin{equation}
\frac{\mathrm{d}}{\mathrm{d}t}b_{\alpha\beta}= \left(v^\gamma{}_{;\alpha} - v^{(n)} b^\gamma{}_{\alpha}\right)b_{\gamma\beta} + \left(b_{\alpha\gamma}v^\gamma\right)_{;\beta} + v^{(n)}_{;\alpha\beta}\text{.}
\end{equation}

Finally we make use of (\ref{eq:contractedGauss2D}) to give $b_{\alpha}{}^{\gamma}b_{\gamma\beta} = 2 H b_{\alpha\beta} -Kg_{\alpha\beta}$ and the symmetry under exchange of indices of $b_{\alpha\beta}$ and the Codazzi equation (\ref{eq:codazzi}) to rewrite $b_{\alpha\gamma;\beta}v^\gamma = v^\gamma b_{\alpha\beta;\gamma}$ and we find
\begin{align}
\frac{\mathrm{d}}{\mathrm{d}t} b_{\alpha\beta} = v^{\gamma}b_{\alpha\beta;\gamma} + b_{\alpha\gamma}v^{\gamma}{}_{;\beta} + b_{\gamma\beta}v^{\gamma}{}_{;\alpha}  + v^{(n)}_{,\alpha;\beta} - v^{(n)}\left(2Hb_{\alpha\beta} - K g_{\alpha\beta}\right)\text{.}
\end{align}

The rate of the area element is given by \citep{Frankel11,Arroyo09}
\begin{equation}
\dot{\mathrm{d}A}= -2 H v^{(n)}\mathrm{d}A\text{.}
\end{equation}

Finally the rate-of-change of the Christoffel symbols is given by \citep{Frankel11}
\begin{equation}
\frac{\mathrm{d}}{\mathrm{d}t} \Gamma^\gamma_{\alpha\beta} = \frac{1}{2}g^{\gamma\delta}\left(\left(\frac{\mathrm{d}}{\mathrm{d}t} g_{\alpha\delta}\right)_{;\beta} + \left(\frac{\mathrm{d}}{\mathrm{d}t} g_{\beta\delta}\right)_{;\alpha} - \left(\frac{\mathrm{d}}{\mathrm{d}t} g_{\alpha\beta}\right)_{;\delta}\right)\text{.}
\end{equation}

\bibliographystyle{jfm}
\bibliography{jfm-instructions}

\end{document}